\documentclass[aps, prl, reprint, groupedaddress, superscriptaddress, showpacs]{revtex4-1}

\pdfoutput=1

\usepackage{graphicx}
\graphicspath{{./figs/}}
\usepackage{mathtools}
\usepackage{commath, amsmath, amssymb, amsfonts}
\usepackage{hyperref}
\usepackage{physics, siunitx}
\usepackage{epstopdf}

\usepackage[normalem]{ulem}
\usepackage[usenames,dvipsnames]{xcolor}

\renewcommand{\va}[1]{\mathbf{#1}}
\newcommand{\field}[1]{\boldsymbol{\mathcal{{#1}}}}
\usepackage{array}
\newcolumntype{P}[1]{>{\centering\arraybackslash}p{#1}}

\usepackage{etoolbox}
\let\bbordermatrix\bordermatrix
\patchcmd{\bbordermatrix}{8.75}{4.75}{}{}
\patchcmd{\bbordermatrix}{\left(}{\left[}{}{}
\patchcmd{\bbordermatrix}{\right)}{\right]}{}{}

\allowdisplaybreaks
\newcommand{\ie}{i.e.~}

\newcommand{\eg}{e.g.~}

\usepackage{mathtools}

\hypersetup{
	colorlinks   = true, 
	urlcolor     = blue, 
	linkcolor    = blue, 
	citecolor   = blue 
}

\begin{document}


\title{Nonlinear excitonic spin Hall effect in monolayer transition metal dichalcogenides}


\author{Alireza Taghizadeh}
\email{ata@nano.aau.dk}
\affiliation{Department of Physics and Nanotechnology, Aalborg University, DK-9220 Aalborg {\O}st, Denmark}


\author{T. G. Pedersen}
\affiliation{Department of Physics and Nanotechnology, Aalborg University, DK-9220 Aalborg {\O}st, Denmark}
\affiliation{Center for Nanostructured Graphene (CNG), DK-9220 Aalborg {\O}st, Denmark}
\date{(Received XX XXXXXX 2018, published XX XXXXXX 2018}%


\date{\today}

\begin{abstract}
We propose and analyze a mechanism for inducing spin Hall currents in ordinary (1H phase) monolayer transition metal dichalcogenides (TMDs) due to the nonlinear process of optical rectification. The photo-induced spin current is proportional to the light intensity, and originates from the intrinsic spin-orbit coupling in TMDs.
The spin current spectrum is strongly influenced by electron-hole interactions, \ie excitonic effects, analogous to the optical absorption. 
Remarkably, excitons change the temperature dependence of the induced spin current, to the point that the current direction can even be reversed by varying the temperature. This peculiar excitonic behavior is shown to emerge from the relative strength of two distinct mechanisms contributing to the optical response, \ie a purely interband part and a mixed inter/intraband contribution. Furthermore, we investigate the valley dichrosim of second-harmonic charge and spin currents, and demonstrate different valley polarization of $s$ and $p$ excitons that stem from their distinct angular momenta. Our findings pave the path to the generation of dc spin currents in ordinary TMDs without external static electric or magnetic fields.
\end{abstract}


\maketitle
The ordinary (charge) Hall effect describes the accumulation of electric charge along the edges of a current-carrying surface in response to an applied magnetic field. A signature of this effect is the emergence of a non-vanishing transverse linear charge conductivity, $\sigma_{xy}^{(1C)}$ \cite{Yoshioka2002}. 
In analogy with the charge Hall effect, the spin Hall effect (SHE) is the spin accumulation at the sample boundaries due to extrinsic \cite{Dyakonov1971, Hirsch1999} or intrinsic \cite{Murakami2003, Sinova2004, Wunderlich2005} spin-orbit coupling (SOC). The SHE was first predicted theoretically \cite{Dyakonov1971, Hirsch1999, Murakami2003} and later observed experimentally in GaAs quantum wells \cite{Kato2004, Wunderlich2005}. In recent years, monolayer transition metal dichalcogenides (TMDs) have been suggested as suitable materials for observing and investigating the SHE, because strong spin-valley coupling enhances the SHE lifetime and hence, eases the experimental observation \cite{Xiao2012, Xu2014}. 
Monolayer TMDs offer large direct bandgaps \cite{Mak2010}, broken crystal inversion symmetry, strong excitonic effects \cite{Qiu2013, Ugeda2014, He2014}, and huge intrinsic SOC \cite{Zhu2011, Xiao2012}, all of which make their optical and electronic properties unique \cite{Mak2016}. In addition, due to the intrinsic coupling of valley and spin degrees of freedom in TMDs, a valley Hall effect coexists simultaneously with the SHE \cite{Xiao2012, Xu2014}. This leads to the valuable possibility of manipulating the valley degree of freedom using the spin in valleytronics \cite{Schaibley2016, Mak2018}. 

\begin{figure}[b!]
	\includegraphics[width=0.49\textwidth]{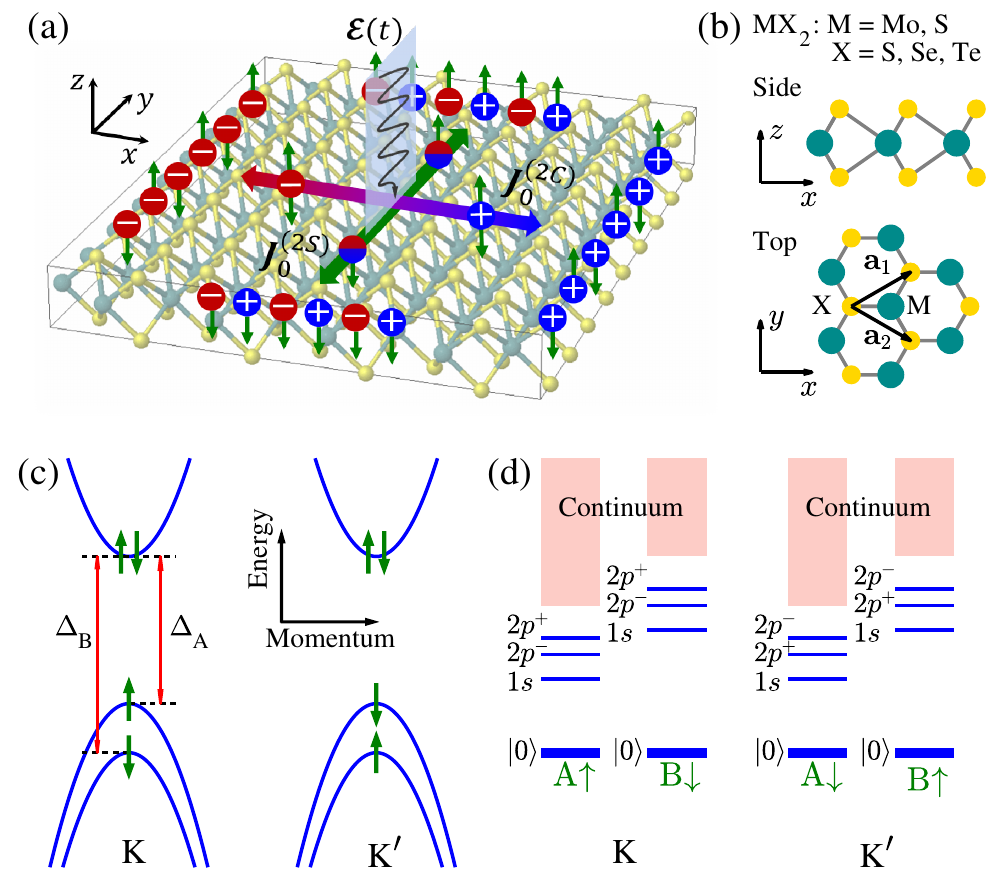}
	\caption[Schematic]{(a) Nonlinear photo-induced Hall effects in monolayer TMDs in response to normal-incident linearly-polarized light. The charge and spin accumulate along perpendicular edges. (b) Top and side view of monolayer TMD (MX$_2$) in 1H phase. (c) Typical quasi-particle band structure at K and K$'$ valleys, where SOC mainly lifts the valence band degeneracy. (d) Sketch of the lowest excitonic states at K and K$'$ valleys for spin-up/down electrons. The A (B) label identifies excitons formed between the conduction band and upper (lower) valence band. }
	\label{fig:Schematic}
\end{figure}

In the absence of a magnetic field, the linear charge Hall effect induced by a time-dependent field in intrinsic monolayer TMDs vanishes due to the time-reversal symmetry (TRS), regardless of the frequency \cite{Ma2018}. In contrast, the linear SHE, characterized by $\sigma_{xy}^{(1S)}$ \cite{Feng2012}, can be non-zero in a specific frequency range, yet with a vanishing static limit (dc) \cite{Feng2012, Pedersen2018}.  
Nonetheless, a finite dc SHE can be obtained in monolayer TMDs with various approaches, \eg by introducing a uniaxial strain to break $C_3$ crystal symmetry \cite{Lee2017}, or by electron/hole doping \cite{Xiao2012}.
Recently, interest has focused on photo-induced Hall effects due to second-order (quadratic) nonlinear processes in non-centrosymmetric materials such as TMDs without breaking TRS \cite{Sodemann2015, Lee2017, Yuan2014, Eginligil2015, Quereda2018, Zhang2018b, Ma2018, Xu2018}. Polarization-dependent charge Hall currents stemming from circular photo-galvanic and photon drag effects have been predicted and observed experimentally in TMDs \cite{Yuan2014, Eginligil2015, Quereda2018}. In addition, quadratic charge Hall currents can be induced in 1T$'$ TMDs that emerge from a non-zero Berry curvature dipole due to reduced symmetry \cite{Sodemann2015, Zhang2018b, Xu2018}. This effect, however, is absent in the usual 1H phase. Apart from being the natural synthesized form of monolayer TMDs, the 1H phase has been widely used for device applications such as transistors \cite{Radisavljevic2011} and photodetectors \cite{Lopez2013, Mak2014}. The alternative phases of monolayer TMDs are either not sufficiently stable \cite{Qian2014, You2018}, \eg the metallic 1T phase, or require strict chemical and thermal conditions \cite{Qian2014}, \eg the topological 1T$'$ phase.
So far, no mechanism for generating dc spin Hall currents in intrinsic 1H monolayer TMDs has been reported. 




In this letter, we identify a quadratic photo-induced spin Hall mechanism leading to a static spin accumulation for monolayer TMDs in the common 1H phase. The induced spin polarization can be detected via spatially-resolved Kerr rotation using a weak optical probe \cite{Kato2004}. 
In Fig.~\ref{fig:Schematic}(a), we depict the Hall geometry to explore the proposed effect, where a normally-incident, linearly-polarized laser beam generates both dc charge and spin Hall currents. 
Under intense irradiation, a field $\field{E}$ oscillating at frequency $\omega$ induces a spin (or charge) dc current at frequency $0=\omega+(-\omega)$ via optical rectification (OR), and simultaneously a fast-oscillating spin (or charge) current at frequency $2\omega=\omega+\omega$ via second-harmonic generation (SHG). The OR and SHG currents are proportional to $\field{E}\field{E}^*$ and $\field{E}^2$, respectively, with frequency-dependent proportionality factors equal to the spin (charge) quadratic optical conductivity tensors $\boldsymbol{\sigma}^{(2S)}$ ($\boldsymbol{\sigma}^{(2C)}$). The 1H crystal structure is invariant under spatial inversion of the $y$-axis ($y\rightarrow -y$) as shown in Fig.~\ref{fig:Schematic}(b), whereas inversion of the $x$-axis ($x\rightarrow -x$) is not a crystal symmetry. This enforces a vanishing $\sigma^{(2C)}_{yyy}$ but a finite $\sigma^{(2C)}_{xxx}$. In contrast, a substantial spin conductivity $\sigma^{(2S)}_{yyy}$ survives in TMDs while $\sigma^{(2S)}_{xxx}$ becomes zero. This surprising property originates from the opposite behavior of the spin compared to charge under time reversal combined with the strong SOC in TMDs. 


Here, we provide a quantum mechanical description of the induced quadratic charge and spin currents including excitonic effects. We demonstrate that both photo-induced charge and spin currents originate from a combination of two different physical mechanisms, \ie inter- and intraband transitions. In monolayer TMDs, the independent-particle interband contribution to the photo-induced charge current vanishes, whereas both inter- and intraband mechanisms contribute to the spin current. Note that the intraband contribution is basically the well-known shift current, which arises from integrating the shift vector over the Brillouin zone \cite{Young2012, Ibanez2018}.
In reality, however, strong excitonic interactions in monolayer TMDs affect both charge and spin currents. In particular, we show that the interband mechanism dominates the induced charge/spin current at high temperature, which can even lead to reversal of the current direction. 
Furthermore, we analyze the dichroic behavior of SHG and the induced quadratic valley polarization. The resonances in SHG spectra display different dichroism, which emerges from the distinct selection rules of $s$ and $p$ excitons. We provide numerical data for MoS$_2$ (here) and WSe$_2$ (Supplementary Material \cite{Supp2018}), but the main findings hold true for other members of the TMD family due to essentially similar physics.

A minimal model for the TMD band structure is obtained with the massive Dirac Hamiltonian \cite{Xiao2012}. 
The Dirac Hamiltonian, which is linear in wavevector $\va{k}$, has proven useful to characterize the electronic properties of TMDs \cite{Xiao2012, Xu2014, Wu2015, Zhou2015, Berkelbach2015, Scharf2017}. However, it fails to account for any even-order nonlinear response (in the dipole approximation) due to the presence of full rotation symmetry, $C_\infty$. By including terms up to second order in $\va{k}$, threefold rotation symmetry, $C_3$, is recovered and trigonal warping (TW) of isoenergy contours is captured \cite{Kormanyos2013}. 
This Hamiltonian, which will be referred to as the TW Hamiltonian hereafter (see \cite{Supp2018}), not only reproduces the band structure more accurately, but also leads to non-vanishing even-order responses.
The typical band structure of monolayer TMDs near the K/K$'$ valleys is shown in Fig.~\ref{fig:Schematic}(c). Note that although the SOC splits both the conduction and valence bands \cite{Xu2014, Wang2018}, the tiny conduction band splitting is ignored here. 

Excitons are known to significantly influence the optical response of monolayer TMDs \cite{Qiu2013, Ugeda2014, He2014, Scharf2017, Wang2018} due to the reduced screening and enhanced confinement of electrons and holes \cite{Pedersen2015, Taghizadeh2018}. The excitonic energies and wave functions can be determined by solving the Bethe-Salpeter equation (BSE) with an appropriate electron-hole interaction kernel \cite{Albrecht1998, Rohlfing2000}. In two-dimensional (2D) materials, the electron-hole interaction is accurately captured by the Keldysh potential \cite{Cudazzo2011}.
Despite the similarities with 2D Hydrogen atoms, the exciton energy spectrum deviates considerably from the hydrogen-like one \cite{Qiu2013, He2014, Wang2015}, which is a signature of Bloch band geometry \cite{Srivastava2015, Zhou2015} and non-local screening \cite{Qiu2013, He2014}.  
The first few states at K and K$'$ for the A and B excitons are shown schematically in Fig.~\ref{fig:Schematic}(d).

The nonlinear conductivity tensors can be determined by solving perturbatively the master equation for the density matrix, $\hat{\rho}(t)$ \cite{Taghizadeh2017, Hipolito2018, Passos2018}. 
To include various scattering mechanisms, we employ a relaxation time approximation \cite{Mermin1970} with two phenomenological broadening parameters, namely $\Gamma_e$ for the coherences ($\rho_{cv\va{k}}$, $\rho_{vc\va{k}}$) and $\Gamma_i$ for the band populations ($\rho_{cc\va{k}}$, $\rho_{vv\va{k}}$). Despite its simplicity, this approximation captures accurately the dynamics of the system \cite{Cheng2015, Passos2018}.
The quadratic charge and spin current densities, $\va{J}^{(2C)}$ and $\va{J}^{(2S)}$, are evaluated as the trace of the density matrix with the corresponding operators \cite{Supp2018}. Without loss of generality, we assume a two-color incident light with the electric field written as $\field{E}(t)= \field{E}_{\omega_1} \exp(-i\omega_1t) + \field{E}_{\omega_2} \exp(-i\omega_2t) + \mathrm{c.c.}$ ($\omega_1$ and $\omega_2$ can be identical). The generated charge and spin current densities at frequency $\omega_1+\omega_2$ are given by $J_{\eta}^{(2X)}(\omega_1+\omega_2)=\sum_{\alpha,\beta} \sigma_{\eta\alpha\beta}^{(2X)}(\omega_1+\omega_2) \mathcal{E}_{\omega_1}^\alpha \mathcal{E}_{\omega_2}^\beta$ with $X=C$/$S$. Note that quadratic charge or spin susceptibilities, $\chi_{\eta\alpha\beta}^{(2X)}$, are related to their corresponding conductivities
by $\epsilon_0 \chi_{\eta\alpha\beta}^{(2X)} \equiv i\sigma_{\eta\alpha\beta}^{(2X)}/(\omega_1+\omega_2)$. The full expressions for charge and spin conductivities are presented in the Supplementary Material \cite{Supp2018}.

Using the TW Hamiltonian, we can distinguish the contributions of spin-up/down electron at the K/K$'$ valleys, which are denoted by $\sigma_{\eta\alpha\beta}^{(2,s\tau)}$. Here, $\tau=\pm1$ and $s=\pm1$ denote the valley and spin indices, respectively. Upon determining $\sigma_{\eta\alpha\beta}^{(2,s\tau)}$, the total charge and spin conductivities read $\sigma_{\eta\alpha\beta}^{(2C)}=\sum_{s,\tau}\sigma_{\eta\alpha\beta}^{(2,s\tau)}$ and $\sigma_{\eta\alpha\beta}^{(2S)}=\sum_{s,\tau} s\sigma_{\eta\alpha\beta}^{(2,s\tau)}$, respectively. Due to the point-group symmetry of the honeycomb lattice, there can be only two independent tensor components of $\sigma_{\eta\alpha\beta}^{(2,s\tau)}$ \cite{Hipolito2016}. In addition, TRS relates the tensor components at the K and K$'$ valleys. The tensor symmetries are summarized in Eq.~(\ref{eq:Symmetries}), in which dots/triangles designate equal magnitudes and open/filled symbols indicate a relative sign difference,
\begin{align}
	\label{eq:Symmetries}
	&\sigma_{\eta\alpha\beta}^{(2,s\tau)}: \hspace{1.4cm} \uparrow\mathrm{K} \hspace{0.5cm} \xLeftrightarrow{\qquad\mathrm{TRS}\qquad} \hspace{0.5cm} \downarrow\mathrm{K}'   \\ 
	&\hspace{0.0cm} \bbordermatrix{\alpha\beta\rightarrow & xx & xy & yx & yy \cr
		\eta\downarrow \quad x & \bullet & \vartriangle & \vartriangle  & \circ \cr
		\qquad \,\,\, y & \vartriangle & \circ & \circ & \blacktriangle \cr} \quad 
	\bbordermatrix{~ & xx & xy & yx & yy \cr
		x & \bullet & \blacktriangle & \blacktriangle  & \circ \cr
		y & \blacktriangle & \circ & \circ & \vartriangle \cr} \nonumber \, .
\end{align}
Using these symmetry relations, it is straightforward to show that $\sigma_{yyy}^{(2C)}=0$ and $\sigma_{xxx}^{(2S)}=0$. 

In the absence of excitonic effects, \ie in the independent particle approximation (IPA), we are able to determine $\sigma_{yyy}^{(2,s\tau)}$ and $\sigma_{xxx}^{(2,s\tau)}$ analytically at zero temperature using the TW Hamiltonian
\begin{subequations}
	\label{eq:IPAConductivity}
	\begin{align}
	\label{eq:IPAConductivityx}
	&\sigma_{xxx}^{(2,s\tau)}(\omega_1+\omega_2) =  \dfrac{i\sqrt{3} e^3 a_0}{48\pi\hbar} \Big[ -\mathcal{G}(\hbar\omega_1+i\Gamma_e,\hbar\omega_2) \nonumber \\
	& \qquad \quad +\mathcal{G}(\hbar\omega_1+\hbar\omega_2+i\Gamma_e,\hbar\omega_2)+(\omega_1\rightleftarrows\omega_2) \Big] \, , \\
	\label{eq:IPAConductivityy}
	&\sigma_{yyy}^{(2,s\tau)}(\omega_1+\omega_2) = \dfrac{\sqrt{3} e^3 a_0}{96 \pi\hbar} \tau \Big[ \mathcal{F}_e(\hbar\omega_1+i\Gamma_e, \hbar\omega_2) \nonumber \\
	& \,\,\, +\mathcal{F}_i(\hbar\omega_1+i\Gamma_e, \hbar\omega_2+i\Gamma_i-i\Gamma_e) + (\omega_1\rightleftarrows\omega_2) \Big] \, . 
	\end{align}
\end{subequations}
Here, $\mathcal{G}(a,b)\equiv\Delta_{s\tau}\tanh[-1](a/\Delta_{s\tau})/(ab)$ with $\Delta_{s\tau}=\{\Delta_\mathrm{A},\Delta_\mathrm{B}\}$ as the effective bandgap depending on the spin and valley indices [see Fig.~\ref{fig:Schematic}(c)], and $(\omega_1\rightleftarrows\omega_2)$ indicates the preceding terms with $\omega_1$ and $\omega_2$ exchanged. The expressions for $\mathcal{F}_e(a,b)$ and $\mathcal{F}_i(a,b)$ are provided in the Supplementary Material \cite{Supp2018} due to their complicated form. 
Note that $\sigma_{xxx}^{(2,s\tau)}$ ($\sigma_{yyy}^{(2,s\tau)}$) has identical (opposite) sign in the two valleys, which is in agreement with TRS. Moreover, $\sigma_{xxx}^{(2,s\tau)}$ includes only intraband transitions, since its interband contribution vanishes, whereas both inter- and intraband transitions contribute to $\sigma_{yyy}^{(2,s\tau)}$. 
Equations~(\ref{eq:IPAConductivityx}) and (\ref{eq:IPAConductivityy}) are valid for any second-order process, \eg SHG: $\omega_1=\omega_2=\omega$; OR: $\omega_1=-\omega_2=\omega$; Pockels effect: $\omega_1=\omega$, $\omega_2=0$. Hereafter, we consider excitation by a monochromatic beam, \ie $\field{E}=\field{E}_\omega\exp(-i\omega t) + \mathrm{c.c.}$, which is either linearly-polarized, \ie $\field{E}_\omega=\mathcal{E}_0\va{e}_y$ or circularly-polarized, \ie $\field{E}_{\omega}=\mathcal{E}_0(\va{e}_x\pm i\va{e}_y)/\sqrt{2}$ ($\va{e}_\alpha$ is the unit vector along the $\alpha$-direction). This generates a quadratic current density $\va{J}^{(2X)}(t)=\va{J}_0^{(2X)}+\va{J}_{2\omega}^{(2X)} \exp(-i2\omega t)  + \mathrm{c.c.}$, where $\va{J}_0^{(2X)}$ and $\va{J}_{2\omega}^{(2X)}$ are the induced OR and SHG current densities, respectively. 




\begin{figure}[t]
	\includegraphics[width=0.49\textwidth]{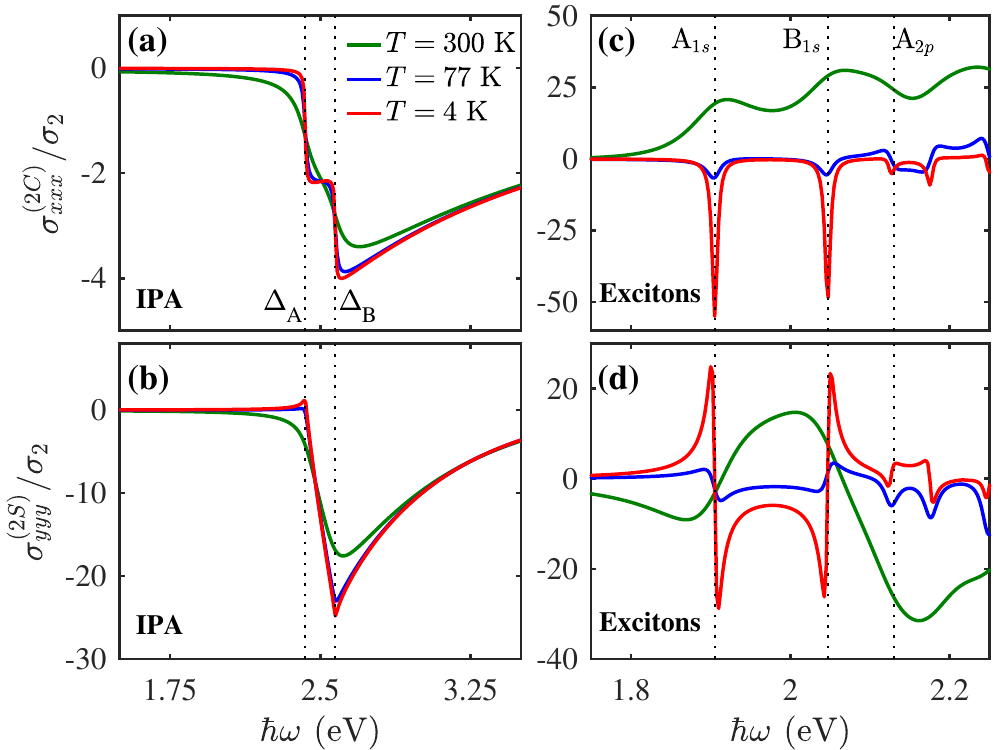}
	\caption[OR Spectrum]{Charge (a,c) and spin (b,d) OR spectra for monolayer MoS$_2$ at $T=300$ K (green), 77 K (blue) and 4 K (red) with the IPA model (a,b) or with excitons (c,d), normalized to $\sigma_2=\SI{1e-15}{SmV^{-1}}$. }
	\label{fig:ORSpectrum}
\end{figure}

\textbf{OR:} Circularly-polarized light at normal incidence generates no dc photo-current in honeycomb lattice crystals due to symmetry constraints \cite{Supp2018}. 
In contrast, linearly-polarized light induces a quadratic charge Hall current in the $x$-direction, $\va{J}_0^{(2C)}=-\sigma_{xxx}^{(2C)}(\omega-\omega)\abs{\mathcal{E}_0}^2\va{e}_x$, and simultaneously a quadratic spin Hall current in the $y$-direction, $\va{J}_0^{(2S)}=\sigma_{yyy}^{(2S)}(\omega-\omega)\abs{\mathcal{E}_0}^2\va{e}_y$. 
Neglecting excitons for the moment, the charge and spin OR conductivities of monolayer MoS$_2$ calculated from Eqs.~(\ref{eq:IPAConductivityx}) and (\ref{eq:IPAConductivityy}) are shown in Figs.~\ref{fig:ORSpectrum}(a) and \ref{fig:ORSpectrum}(b), respectively. Focusing on the general features of the IPA response, $\sigma_{xxx}^{(2C)}$ shows two steps appearing at $\hbar\omega=\{\Delta_\mathrm{A},\Delta_\mathrm{B}\}$, that emerge from contributions of spin-up and spin-down electrons, respectively. In contrast, $\sigma_{yyy}^{(2S)}$ exhibits a jump at $\hbar\omega=\Delta_\mathrm{A}$ followed by a jump in the opposite direction at $\hbar\omega=\Delta_\mathrm{B}$.
For both $\sigma_{xxx}^{(2C)}$ and $\sigma_{yyy}^{(2S)}$, the response at large frequencies ($\hbar\omega>\Delta_\mathrm{B}$) decreases approximately as $1/\omega$. It is interesting to note that the spin conductivity is one order of magnitude larger than the charge response.

Next, we discuss the differences between the three sets of curves in Fig.~\ref{fig:ORSpectrum}, which are obtained for three different temperatures $T$. Changing $T$ affects various parameters such as the bandgap, hopping (effective mass), equilibrium Fermi distribution, and broadenings. Among these parameters, the variation of hopping and $\Gamma_i$ is negligible, whereas the Fermi distribution only affects the optical response marginally in undoped monolayer TMDs, since the bandgap is much larger than $k_BT$. The bandgap shrinks approximately 5\% if $T$ is raised from 4 K to 300 K \cite{Cadiz2017}, and thereby simply red-shifts the entire spectrum. To ease comparison, this shift is not included in the plots. In contrast, the interband broadening, $\Gamma_e$, varies significantly with temperature, and leads to strong modifications of the spectra \cite{Selig2016, Christiansen2017, Scuri2018}. 
The different temperature behaviors of $\Gamma_i$ and $\Gamma_e$ follow from their distinct origin. Physically speaking, $\Gamma_i$ originates mainly from carrier-carrier (electron or hole) scattering, and is closely related to the Drude response in doped TMDs \cite{Shen2013, Gupta2018}. In contrast, $\Gamma_e$ includes various phonon-assisted, impurity/defect-related and pure dephasing scattering mechanisms \cite{Scuri2018}. While the disorder-related contributions are nearly independent of the temperature, the pure dephasing and phonon-assisted parts increase with $T$ due to enhanced carrier-phonon scattering \cite{Selig2016, Christiansen2017, Scuri2018}. 
For monolayer MoS$_2$, $\Gamma_i$ is measured to be approximately 25 meV \cite{Scuri2018}, whereas $\Gamma_e$ is estimated to be 4, 10, 53 meV \cite{Cadiz2017} at $T=4$, 77, 300 K, respectively.
Hence, in the IPA result of Figs.~\ref{fig:ORSpectrum}(a) and \ref{fig:ORSpectrum}(b), smaller values of $\Gamma_e$ at lower temperatures mean sharper spectral features.




Including excitonic effects, the charge and spin OR conductivities are illustrated in Figs.~\ref{fig:ORSpectrum}(c) and \ref{fig:ORSpectrum}(d), respectively. In contrast to the IPA results, the nonlinear conductivities exhibit discrete resonances inside the bandgap as typically observed for excitonic optical responses \cite{Pedersen2015, Taghizadeh2018}. Here, the first three resonances at 1.9, 2.05, 2.15 eV are labeled according to their characteristic wave function symmetry as $\mathrm{A}_{1s}$, $\mathrm{B}_{1s}$ and $\mathrm{A}_{2p}$, respectively \cite{Ye2014}. In contrast to the linear optical response, where mainly $s$ excitons are bright \cite{Cao2018, Zhang2018}, both $s$ and $p$ excitons manifest themselves in the quadratic optical response \cite{Ye2014, Wang2015}. Regarding the temperature-dependence, a significant difference from the IPA result is observed: {\it increasing the temperature not only broadens the spectral features, but also inverts the sign of the conductivity}. Hence, changing the temperature may reverse the direction of charge or spin currents. This prominent and peculiar behavior is a purely excitonic effect that stems from the relative change of inter- and intraband contributions \cite{Supp2018}. At low $T$, the response is dominated by contributions that arise from coherence terms. By increasing the temperature, exciton-phonon scattering is enhanced significantly, which leads to larger change of band populations with respect to their equilibrium values. Thus, the relative weight of band population contributions is increased compared to coherence terms.


\begin{figure}[t]
	\includegraphics[width=0.49\textwidth]{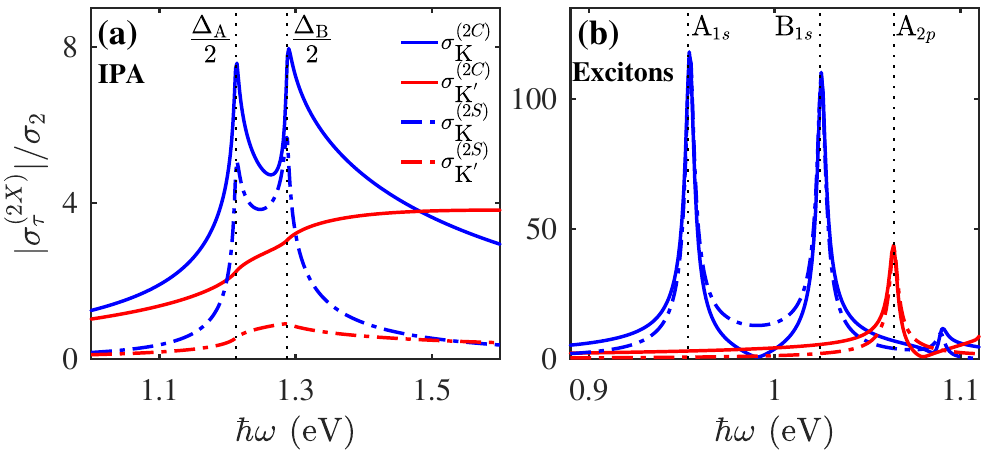}
	\caption[SHG Spectrum]{Contribution of K (blue) and K$'$ (red) valleys to charge (solid) and spin (dashed) SHG conductivities for monolayer MoS$_2$ at $T=4$ K with the IPA model (a) or excitons (b) in response to a left circularly-polarized beam ($\sigma_2=\SI{1e-15}{SmV^{-1}}$).} 
	\label{fig:SHGSpectrum}
\end{figure}


\textbf{SHG:} In addition to the generated dc photo-current, second-harmonic spin and charge currents, $\va{J}_{2\omega}^{(2X)}$, are induced in monolayer TMDs. In the case of linearly-polarized light, the properties of these fast-oscillating charge and spin currents have already been discussed in Ref.~\onlinecite{Pedersen2018}. Hence, in the present work, we only focus on the induced SHG optical response obtained for circularly-polarized excitation. For left-hand circular polarization, the second-harmonic current density is given by $\va{J}_{2\omega}^{(2C)}=\sigma_{xxx}^{(2C)}(\omega+\omega)\mathcal{E}_0^2(\va{e}_x-i\va{e}_y)$, whereas its spin counterpart becomes $\va{J}_{2\omega}^{(2S)}=-\sigma_{yyy}^{(2S)}(\omega+\omega)\mathcal{E}_0^2(\va{e}_x+i\va{e}_y)$ \cite{Supp2018}. The induced charge current (or equivalently polarization) rotates in the opposite direction compared to the incident electric field, whereas the spin response rotates in the same direction. In contrast to the SHG charge current, which can readily be detected due to its electromagnetic radiation \cite{Ye2014, Seyler2015}, the detection of SHG spin current requires advanced coherent techniques \cite{Zhao2006, Wang2010, Werake2010}. 

It is well known that a circularly-polarized beam can excite an individual valley in one-photon processes such as absorption \cite{Zeng2012, Mak2012, Cao2012}. Similarly, for two-photon processes such as SHG, the generated current is valley-dependent.
This nonlinear valley-contrasting effect can be characterized by defining an effective conductivity for each individual valley in response to circularly-polarized light, \ie $\sigma_{xxx}^{(2C)}(\omega+\omega)=\sigma_{\mathrm{K}}^{(2C)}+\sigma_{\mathrm{K}'}^{(2C)}$ and $\sigma_{yyy}^{(2S)}(\omega+\omega)=\sigma_{\mathrm{K}}^{(2S)}+\sigma_{\mathrm{K}'}^{(2S)}$ \cite{Supp2018}.
The valley conductivities, obtained in response to left-circular light for monolayer MoS$_2$ with the IPA model, are presented in Fig.~\ref{fig:SHGSpectrum}(a). For both charge and spin conductivities, although the K valley contributes more than the K$'$ valley to the generated current, their contributions are of comparable magnitude. The valley dichroism is enhanced significantly when excitonic effects are included as shown in Fig.~\ref{fig:SHGSpectrum}(b). Here, the K valley dominates the charge or spin currents due to $s$ excitons ($\mathrm{A}_{1s}$ and $\mathrm{B}_{1s}$), whereas the response of $p$ excitons emerges mainly from the K$'$ valley. This different behavior of $s$ and $p$ excitons stems from distinct nonlinear selection rules due to their angular momenta \cite{Xiao2015}.
For right circularly-polarized excitation, the K and K$'$ valleys switch their roles. The valley-contrasting physics appearing in the nonlinear response originates from the crystal point symmetry, in analogy with the linear response \cite{Zeng2012, Mak2012, Cao2012}.



In conclusion, we have shown that a nonlinear dc spin Hall current is induced by linearly-polarized light in intrinsic 1H monolayer TMDs. The spin current, that stems from the large SOC, does not require any strain, external static electric or magnetic field. Similarly to the linear optical response, excitons significantly modify the quadratic optical response. We predict that the spin current direction can be inverted, by varying the temperature exclusively due to excitonic effects. Finally, the distinct valley-dependence of the induced second-harmonic current under circularly-polarized light for $s$ and $p$ excitons is demonstrated. 

\acknowledgments
The authors thank F. Hipolito for helpful discussions throughout the project. 
This work was supported by the QUSCOPE center sponsored by the Villum Foundation and TGP is financially supported by the CNG center under the Danish National Research Foundation, project DNRF103.

\bibliography{SpinHall}

\begin{thebibliography}{71}%
\makeatletter
\providecommand \@ifxundefined [1]{%
 \@ifx{#1\undefined}
}%
\providecommand \@ifnum [1]{%
 \ifnum #1\expandafter \@firstoftwo
 \else \expandafter \@secondoftwo
 \fi
}%
\providecommand \@ifx [1]{%
 \ifx #1\expandafter \@firstoftwo
 \else \expandafter \@secondoftwo
 \fi
}%
\providecommand \natexlab [1]{#1}%
\providecommand \enquote  [1]{``#1''}%
\providecommand \bibnamefont  [1]{#1}%
\providecommand \bibfnamefont [1]{#1}%
\providecommand \citenamefont [1]{#1}%
\providecommand \href@noop [0]{\@secondoftwo}%
\providecommand \href [0]{\begingroup \@sanitize@url \@href}%
\providecommand \@href[1]{\@@startlink{#1}\@@href}%
\providecommand \@@href[1]{\endgroup#1\@@endlink}%
\providecommand \@sanitize@url [0]{\catcode `\\12\catcode `\$12\catcode
  `\&12\catcode `\#12\catcode `\^12\catcode `\_12\catcode `\%12\relax}%
\providecommand \@@startlink[1]{}%
\providecommand \@@endlink[0]{}%
\providecommand \url  [0]{\begingroup\@sanitize@url \@url }%
\providecommand \@url [1]{\endgroup\@href {#1}{\urlprefix }}%
\providecommand \urlprefix  [0]{URL }%
\providecommand \Eprint [0]{\href }%
\providecommand \doibase [0]{http://dx.doi.org/}%
\providecommand \selectlanguage [0]{\@gobble}%
\providecommand \bibinfo  [0]{\@secondoftwo}%
\providecommand \bibfield  [0]{\@secondoftwo}%
\providecommand \translation [1]{[#1]}%
\providecommand \BibitemOpen [0]{}%
\providecommand \bibitemStop [0]{}%
\providecommand \bibitemNoStop [0]{.\EOS\space}%
\providecommand \EOS [0]{\spacefactor3000\relax}%
\providecommand \BibitemShut  [1]{\csname bibitem#1\endcsname}%
\let\auto@bib@innerbib\@empty
\bibitem [{\citenamefont {Yoshioka}(2002)}]{Yoshioka2002}%
  \BibitemOpen
  \bibfield  {author} {\bibinfo {author} {\bibfnamefont {D.}~\bibnamefont
  {Yoshioka}},\ }\href {\doibase 10.1007/978-3-662-05016-3} {\emph {\bibinfo
  {title} {The Quantum Hall Effect}}}\ (\bibinfo  {publisher} {Springer Berlin
  Heidelberg},\ \bibinfo {year} {2002})\BibitemShut {NoStop}%
\bibitem [{\citenamefont {Dyakonov}\ and\ \citenamefont
  {Perel}(1971)}]{Dyakonov1971}%
  \BibitemOpen
  \bibfield  {author} {\bibinfo {author} {\bibfnamefont {M.}~\bibnamefont
  {Dyakonov}}\ and\ \bibinfo {author} {\bibfnamefont {V.}~\bibnamefont
  {Perel}},\ }\href {\doibase 10.1016/0375-9601(71)90196-4} {\bibfield
  {journal} {\bibinfo  {journal} {Phys. Lett. A}\ }\textbf {\bibinfo {volume}
  {35}},\ \bibinfo {pages} {459} (\bibinfo {year} {1971})}\BibitemShut
  {NoStop}%
\bibitem [{\citenamefont {Hirsch}(1999)}]{Hirsch1999}%
  \BibitemOpen
  \bibfield  {author} {\bibinfo {author} {\bibfnamefont {J.~E.}\ \bibnamefont
  {Hirsch}},\ }\href {\doibase 10.1103/PhysRevLett.83.1834} {\bibfield
  {journal} {\bibinfo  {journal} {Phys. Rev. Lett.}\ }\textbf {\bibinfo
  {volume} {83}},\ \bibinfo {pages} {1834} (\bibinfo {year}
  {1999})}\BibitemShut {NoStop}%
\bibitem [{\citenamefont {Murakami}(2003)}]{Murakami2003}%
  \BibitemOpen
  \bibfield  {author} {\bibinfo {author} {\bibfnamefont {S.}~\bibnamefont
  {Murakami}},\ }\href {\doibase 10.1126/science.1087128} {\bibfield  {journal}
  {\bibinfo  {journal} {Science}\ }\textbf {\bibinfo {volume} {301}},\ \bibinfo
  {pages} {1348} (\bibinfo {year} {2003})}\BibitemShut {NoStop}%
\bibitem [{\citenamefont {Sinova}\ \emph {et~al.}(2004)\citenamefont {Sinova},
  \citenamefont {Culcer}, \citenamefont {Niu}, \citenamefont {Sinitsyn},
  \citenamefont {Jungwirth},\ and\ \citenamefont {MacDonald}}]{Sinova2004}%
  \BibitemOpen
  \bibfield  {author} {\bibinfo {author} {\bibfnamefont {J.}~\bibnamefont
  {Sinova}}, \bibinfo {author} {\bibfnamefont {D.}~\bibnamefont {Culcer}},
  \bibinfo {author} {\bibfnamefont {Q.}~\bibnamefont {Niu}}, \bibinfo {author}
  {\bibfnamefont {N.~A.}\ \bibnamefont {Sinitsyn}}, \bibinfo {author}
  {\bibfnamefont {T.}~\bibnamefont {Jungwirth}}, \ and\ \bibinfo {author}
  {\bibfnamefont {A.~H.}\ \bibnamefont {MacDonald}},\ }\href {\doibase
  10.1103/PhysRevLett.92.126603} {\bibfield  {journal} {\bibinfo  {journal}
  {Phys. Rev. Lett.}\ }\textbf {\bibinfo {volume} {92}},\ \bibinfo {pages}
  {126603} (\bibinfo {year} {2004})}\BibitemShut {NoStop}%
\bibitem [{\citenamefont {Wunderlich}\ \emph {et~al.}(2005)\citenamefont
  {Wunderlich}, \citenamefont {Kaestner}, \citenamefont {Sinova},\ and\
  \citenamefont {Jungwirth}}]{Wunderlich2005}%
  \BibitemOpen
  \bibfield  {author} {\bibinfo {author} {\bibfnamefont {J.}~\bibnamefont
  {Wunderlich}}, \bibinfo {author} {\bibfnamefont {B.}~\bibnamefont
  {Kaestner}}, \bibinfo {author} {\bibfnamefont {J.}~\bibnamefont {Sinova}}, \
  and\ \bibinfo {author} {\bibfnamefont {T.}~\bibnamefont {Jungwirth}},\ }\href
  {\doibase 10.1103/PhysRevLett.94.047204} {\bibfield  {journal} {\bibinfo
  {journal} {Phys. Rev. Lett.}\ }\textbf {\bibinfo {volume} {94}},\ \bibinfo
  {pages} {047204} (\bibinfo {year} {2005})}\BibitemShut {NoStop}%
\bibitem [{\citenamefont {Kato}\ \emph {et~al.}(2004)\citenamefont {Kato},
  \citenamefont {Myers}, \citenamefont {Gossard},\ and\ \citenamefont
  {Awschalom}}]{Kato2004}%
  \BibitemOpen
  \bibfield  {author} {\bibinfo {author} {\bibfnamefont {Y.~K.}\ \bibnamefont
  {Kato}}, \bibinfo {author} {\bibfnamefont {R.~C.}\ \bibnamefont {Myers}},
  \bibinfo {author} {\bibfnamefont {A.~C.}\ \bibnamefont {Gossard}}, \ and\
  \bibinfo {author} {\bibfnamefont {D.~D.}\ \bibnamefont {Awschalom}},\ }\href
  {\doibase 10.1126/science.1105514} {\bibfield  {journal} {\bibinfo  {journal}
  {Science}\ }\textbf {\bibinfo {volume} {306}},\ \bibinfo {pages} {1910}
  (\bibinfo {year} {2004})}\BibitemShut {NoStop}%
\bibitem [{\citenamefont {Xiao}\ \emph {et~al.}(2012)\citenamefont {Xiao},
  \citenamefont {Liu}, \citenamefont {Feng}, \citenamefont {Xu},\ and\
  \citenamefont {Yao}}]{Xiao2012}%
  \BibitemOpen
  \bibfield  {author} {\bibinfo {author} {\bibfnamefont {D.}~\bibnamefont
  {Xiao}}, \bibinfo {author} {\bibfnamefont {G.-B.}\ \bibnamefont {Liu}},
  \bibinfo {author} {\bibfnamefont {W.}~\bibnamefont {Feng}}, \bibinfo {author}
  {\bibfnamefont {X.}~\bibnamefont {Xu}}, \ and\ \bibinfo {author}
  {\bibfnamefont {W.}~\bibnamefont {Yao}},\ }\href {\doibase
  10.1103/PhysRevLett.108.196802} {\bibfield  {journal} {\bibinfo  {journal}
  {Phys. Rev. Lett.}\ }\textbf {\bibinfo {volume} {108}},\ \bibinfo {pages}
  {196802} (\bibinfo {year} {2012})}\BibitemShut {NoStop}%
\bibitem [{\citenamefont {Xu}\ \emph {et~al.}(2014)\citenamefont {Xu},
  \citenamefont {Yao}, \citenamefont {Xiao},\ and\ \citenamefont
  {Heinz}}]{Xu2014}%
  \BibitemOpen
  \bibfield  {author} {\bibinfo {author} {\bibfnamefont {X.}~\bibnamefont
  {Xu}}, \bibinfo {author} {\bibfnamefont {W.}~\bibnamefont {Yao}}, \bibinfo
  {author} {\bibfnamefont {D.}~\bibnamefont {Xiao}}, \ and\ \bibinfo {author}
  {\bibfnamefont {T.~F.}\ \bibnamefont {Heinz}},\ }\href {\doibase
  10.1038/nphys2942} {\bibfield  {journal} {\bibinfo  {journal} {Nat. Phys.}\
  }\textbf {\bibinfo {volume} {10}},\ \bibinfo {pages} {343} (\bibinfo {year}
  {2014})}\BibitemShut {NoStop}%
\bibitem [{\citenamefont {Mak}\ \emph {et~al.}(2010)\citenamefont {Mak},
  \citenamefont {Lee}, \citenamefont {Hone}, \citenamefont {Shan},\ and\
  \citenamefont {Heinz}}]{Mak2010}%
  \BibitemOpen
  \bibfield  {author} {\bibinfo {author} {\bibfnamefont {K.~F.}\ \bibnamefont
  {Mak}}, \bibinfo {author} {\bibfnamefont {C.}~\bibnamefont {Lee}}, \bibinfo
  {author} {\bibfnamefont {J.}~\bibnamefont {Hone}}, \bibinfo {author}
  {\bibfnamefont {J.}~\bibnamefont {Shan}}, \ and\ \bibinfo {author}
  {\bibfnamefont {T.~F.}\ \bibnamefont {Heinz}},\ }\href {\doibase
  10.1103/PhysRevLett.105.136805} {\bibfield  {journal} {\bibinfo  {journal}
  {Phys. Rev. Lett.}\ }\textbf {\bibinfo {volume} {105}},\ \bibinfo {pages}
  {136805} (\bibinfo {year} {2010})}\BibitemShut {NoStop}%
\bibitem [{\citenamefont {Qiu}\ \emph {et~al.}(2013)\citenamefont {Qiu},
  \citenamefont {da~Jornada},\ and\ \citenamefont {Louie}}]{Qiu2013}%
  \BibitemOpen
  \bibfield  {author} {\bibinfo {author} {\bibfnamefont {D.~Y.}\ \bibnamefont
  {Qiu}}, \bibinfo {author} {\bibfnamefont {F.~H.}\ \bibnamefont {da~Jornada}},
  \ and\ \bibinfo {author} {\bibfnamefont {S.~G.}\ \bibnamefont {Louie}},\
  }\href {\doibase 10.1103/PhysRevLett.111.216805} {\bibfield  {journal}
  {\bibinfo  {journal} {Phys. Rev. Lett.}\ }\textbf {\bibinfo {volume} {111}},\
  \bibinfo {pages} {216805} (\bibinfo {year} {2013})}\BibitemShut {NoStop}%
\bibitem [{\citenamefont {Ugeda}\ \emph {et~al.}(2014)\citenamefont {Ugeda},
  \citenamefont {Bradley}, \citenamefont {Shi}, \citenamefont {da~Jornada},
  \citenamefont {Zhang}, \citenamefont {Qiu}, \citenamefont {Ruan},
  \citenamefont {Mo}, \citenamefont {Hussain}, \citenamefont {Shen},
  \citenamefont {Wang}, \citenamefont {Louie},\ and\ \citenamefont
  {Crommie}}]{Ugeda2014}%
  \BibitemOpen
  \bibfield  {author} {\bibinfo {author} {\bibfnamefont {M.~M.}\ \bibnamefont
  {Ugeda}}, \bibinfo {author} {\bibfnamefont {A.~J.}\ \bibnamefont {Bradley}},
  \bibinfo {author} {\bibfnamefont {S.-F.}\ \bibnamefont {Shi}}, \bibinfo
  {author} {\bibfnamefont {F.~H.}\ \bibnamefont {da~Jornada}}, \bibinfo
  {author} {\bibfnamefont {Y.}~\bibnamefont {Zhang}}, \bibinfo {author}
  {\bibfnamefont {D.~Y.}\ \bibnamefont {Qiu}}, \bibinfo {author} {\bibfnamefont
  {W.}~\bibnamefont {Ruan}}, \bibinfo {author} {\bibfnamefont {S.-K.}\
  \bibnamefont {Mo}}, \bibinfo {author} {\bibfnamefont {Z.}~\bibnamefont
  {Hussain}}, \bibinfo {author} {\bibfnamefont {Z.-X.}\ \bibnamefont {Shen}},
  \bibinfo {author} {\bibfnamefont {F.}~\bibnamefont {Wang}}, \bibinfo {author}
  {\bibfnamefont {S.~G.}\ \bibnamefont {Louie}}, \ and\ \bibinfo {author}
  {\bibfnamefont {M.~F.}\ \bibnamefont {Crommie}},\ }\href {\doibase
  10.1038/nmat4061} {\bibfield  {journal} {\bibinfo  {journal} {Nat. Mater.}\
  }\textbf {\bibinfo {volume} {13}},\ \bibinfo {pages} {1091} (\bibinfo {year}
  {2014})}\BibitemShut {NoStop}%
\bibitem [{\citenamefont {He}\ \emph {et~al.}(2014)\citenamefont {He},
  \citenamefont {Kumar}, \citenamefont {Zhao}, \citenamefont {Wang},
  \citenamefont {Mak}, \citenamefont {Zhao},\ and\ \citenamefont
  {Shan}}]{He2014}%
  \BibitemOpen
  \bibfield  {author} {\bibinfo {author} {\bibfnamefont {K.}~\bibnamefont
  {He}}, \bibinfo {author} {\bibfnamefont {N.}~\bibnamefont {Kumar}}, \bibinfo
  {author} {\bibfnamefont {L.}~\bibnamefont {Zhao}}, \bibinfo {author}
  {\bibfnamefont {Z.}~\bibnamefont {Wang}}, \bibinfo {author} {\bibfnamefont
  {K.~F.}\ \bibnamefont {Mak}}, \bibinfo {author} {\bibfnamefont
  {H.}~\bibnamefont {Zhao}}, \ and\ \bibinfo {author} {\bibfnamefont
  {J.}~\bibnamefont {Shan}},\ }\href {\doibase 10.1103/PhysRevLett.113.026803}
  {\bibfield  {journal} {\bibinfo  {journal} {Phys. Rev. Lett.}\ }\textbf
  {\bibinfo {volume} {113}},\ \bibinfo {pages} {026803} (\bibinfo {year}
  {2014})}\BibitemShut {NoStop}%
\bibitem [{\citenamefont {Zhu}\ \emph {et~al.}(2011)\citenamefont {Zhu},
  \citenamefont {Cheng},\ and\ \citenamefont {Schwingenschl\"ogl}}]{Zhu2011}%
  \BibitemOpen
  \bibfield  {author} {\bibinfo {author} {\bibfnamefont {Z.~Y.}\ \bibnamefont
  {Zhu}}, \bibinfo {author} {\bibfnamefont {Y.~C.}\ \bibnamefont {Cheng}}, \
  and\ \bibinfo {author} {\bibfnamefont {U.}~\bibnamefont
  {Schwingenschl\"ogl}},\ }\href {\doibase 10.1103/PhysRevB.84.153402}
  {\bibfield  {journal} {\bibinfo  {journal} {Phys. Rev. B}\ }\textbf {\bibinfo
  {volume} {84}},\ \bibinfo {pages} {153402} (\bibinfo {year}
  {2011})}\BibitemShut {NoStop}%
\bibitem [{\citenamefont {Mak}\ and\ \citenamefont {Shan}(2016)}]{Mak2016}%
  \BibitemOpen
  \bibfield  {author} {\bibinfo {author} {\bibfnamefont {K.~F.}\ \bibnamefont
  {Mak}}\ and\ \bibinfo {author} {\bibfnamefont {J.}~\bibnamefont {Shan}},\
  }\href {\doibase 10.1038/nphoton.2015.282} {\bibfield  {journal} {\bibinfo
  {journal} {Nat. Photon.}\ }\textbf {\bibinfo {volume} {10}},\ \bibinfo
  {pages} {216} (\bibinfo {year} {2016})}\BibitemShut {NoStop}%
\bibitem [{\citenamefont {Schaibley}\ \emph {et~al.}(2016)\citenamefont
  {Schaibley}, \citenamefont {Yu}, \citenamefont {Clark}, \citenamefont
  {Rivera}, \citenamefont {Ross}, \citenamefont {Seyler}, \citenamefont {Yao},\
  and\ \citenamefont {Xu}}]{Schaibley2016}%
  \BibitemOpen
  \bibfield  {author} {\bibinfo {author} {\bibfnamefont {J.~R.}\ \bibnamefont
  {Schaibley}}, \bibinfo {author} {\bibfnamefont {H.}~\bibnamefont {Yu}},
  \bibinfo {author} {\bibfnamefont {G.}~\bibnamefont {Clark}}, \bibinfo
  {author} {\bibfnamefont {P.}~\bibnamefont {Rivera}}, \bibinfo {author}
  {\bibfnamefont {J.~S.}\ \bibnamefont {Ross}}, \bibinfo {author}
  {\bibfnamefont {K.~L.}\ \bibnamefont {Seyler}}, \bibinfo {author}
  {\bibfnamefont {W.}~\bibnamefont {Yao}}, \ and\ \bibinfo {author}
  {\bibfnamefont {X.}~\bibnamefont {Xu}},\ }\href {\doibase
  10.1038/natrevmats.2016.55} {\bibfield  {journal} {\bibinfo  {journal} {Nat.
  Rev. Mater.}\ }\textbf {\bibinfo {volume} {1}},\ \bibinfo {pages} {16055}
  (\bibinfo {year} {2016})}\BibitemShut {NoStop}%
\bibitem [{\citenamefont {Mak}\ \emph {et~al.}(2018)\citenamefont {Mak},
  \citenamefont {Xiao},\ and\ \citenamefont {Shan}}]{Mak2018}%
  \BibitemOpen
  \bibfield  {author} {\bibinfo {author} {\bibfnamefont {K.~F.}\ \bibnamefont
  {Mak}}, \bibinfo {author} {\bibfnamefont {D.}~\bibnamefont {Xiao}}, \ and\
  \bibinfo {author} {\bibfnamefont {J.}~\bibnamefont {Shan}},\ }\href {\doibase
  10.1038/s41566-018-0204-6} {\bibfield  {journal} {\bibinfo  {journal} {Nat.
  Photon.}\ }\textbf {\bibinfo {volume} {12}},\ \bibinfo {pages} {451}
  (\bibinfo {year} {2018})}\BibitemShut {NoStop}%
\bibitem [{\citenamefont {Ma}\ \emph {et~al.}()\citenamefont {Ma},
  \citenamefont {Xu}, \citenamefont {Shen}, \citenamefont {Macneill},
  \citenamefont {Fatemi}, \citenamefont {Valdivia}, \citenamefont {Wu},
  \citenamefont {Chang}, \citenamefont {Du}, \citenamefont {Hsu}, \citenamefont
  {Gibson}, \citenamefont {Fang}, \citenamefont {Kaxiras}, \citenamefont
  {Watanabe}, \citenamefont {Taniguchi}, \citenamefont {Cava}, \citenamefont
  {Lu}, \citenamefont {Lin}, \citenamefont {Fu}, \citenamefont {Gedik},\ and\
  \citenamefont {Jarillo-Herrero}}]{Ma2018}%
  \BibitemOpen
  \bibfield  {author} {\bibinfo {author} {\bibfnamefont {Q.}~\bibnamefont
  {Ma}}, \bibinfo {author} {\bibfnamefont {S.-Y.}\ \bibnamefont {Xu}}, \bibinfo
  {author} {\bibfnamefont {H.}~\bibnamefont {Shen}}, \bibinfo {author}
  {\bibfnamefont {D.}~\bibnamefont {Macneill}}, \bibinfo {author}
  {\bibfnamefont {V.}~\bibnamefont {Fatemi}}, \bibinfo {author} {\bibfnamefont
  {A.~M.~M.}\ \bibnamefont {Valdivia}}, \bibinfo {author} {\bibfnamefont
  {S.}~\bibnamefont {Wu}}, \bibinfo {author} {\bibfnamefont {T.-R.}\
  \bibnamefont {Chang}}, \bibinfo {author} {\bibfnamefont {Z.}~\bibnamefont
  {Du}}, \bibinfo {author} {\bibfnamefont {C.-H.}\ \bibnamefont {Hsu}},
  \bibinfo {author} {\bibfnamefont {Q.~D.}\ \bibnamefont {Gibson}}, \bibinfo
  {author} {\bibfnamefont {S.}~\bibnamefont {Fang}}, \bibinfo {author}
  {\bibfnamefont {E.}~\bibnamefont {Kaxiras}}, \bibinfo {author} {\bibfnamefont
  {K.}~\bibnamefont {Watanabe}}, \bibinfo {author} {\bibfnamefont
  {T.}~\bibnamefont {Taniguchi}}, \bibinfo {author} {\bibfnamefont {R.~J.}\
  \bibnamefont {Cava}}, \bibinfo {author} {\bibfnamefont {H.-Z.}\ \bibnamefont
  {Lu}}, \bibinfo {author} {\bibfnamefont {H.}~\bibnamefont {Lin}}, \bibinfo
  {author} {\bibfnamefont {L.}~\bibnamefont {Fu}}, \bibinfo {author}
  {\bibfnamefont {N.}~\bibnamefont {Gedik}}, \ and\ \bibinfo {author}
  {\bibfnamefont {P.}~\bibnamefont {Jarillo-Herrero}},\ }\href
  {https://arxiv.org/abs/1809.09279} {\ }\Eprint
  {http://arxiv.org/abs/1809.09279} {arXiv:1809.09279} \BibitemShut {NoStop}%
\bibitem [{\citenamefont {Feng}\ \emph {et~al.}(2012)\citenamefont {Feng},
  \citenamefont {Yao}, \citenamefont {Zhu}, \citenamefont {Zhou}, \citenamefont
  {Yao},\ and\ \citenamefont {Xiao}}]{Feng2012}%
  \BibitemOpen
  \bibfield  {author} {\bibinfo {author} {\bibfnamefont {W.}~\bibnamefont
  {Feng}}, \bibinfo {author} {\bibfnamefont {Y.}~\bibnamefont {Yao}}, \bibinfo
  {author} {\bibfnamefont {W.}~\bibnamefont {Zhu}}, \bibinfo {author}
  {\bibfnamefont {J.}~\bibnamefont {Zhou}}, \bibinfo {author} {\bibfnamefont
  {W.}~\bibnamefont {Yao}}, \ and\ \bibinfo {author} {\bibfnamefont
  {D.}~\bibnamefont {Xiao}},\ }\href {\doibase 10.1103/PhysRevB.86.165108}
  {\bibfield  {journal} {\bibinfo  {journal} {Phys. Rev. B}\ }\textbf {\bibinfo
  {volume} {86}},\ \bibinfo {pages} {165108} (\bibinfo {year}
  {2012})}\BibitemShut {NoStop}%
\bibitem [{\citenamefont {Pedersen}(2018)}]{Pedersen2018}%
  \BibitemOpen
  \bibfield  {author} {\bibinfo {author} {\bibfnamefont {T.~G.}\ \bibnamefont
  {Pedersen}},\ }\href {\doibase 10.1103/PhysRevB.98.165425} {\bibfield
  {journal} {\bibinfo  {journal} {Phys. Rev. B}\ }\textbf {\bibinfo {volume}
  {98}},\ \bibinfo {pages} {165425} (\bibinfo {year} {2018})}\BibitemShut
  {NoStop}%
\bibitem [{\citenamefont {Lee}\ \emph {et~al.}(2017)\citenamefont {Lee},
  \citenamefont {Wang}, \citenamefont {Xie}, \citenamefont {Mak},\ and\
  \citenamefont {Shan}}]{Lee2017}%
  \BibitemOpen
  \bibfield  {author} {\bibinfo {author} {\bibfnamefont {J.}~\bibnamefont
  {Lee}}, \bibinfo {author} {\bibfnamefont {Z.}~\bibnamefont {Wang}}, \bibinfo
  {author} {\bibfnamefont {H.}~\bibnamefont {Xie}}, \bibinfo {author}
  {\bibfnamefont {K.~F.}\ \bibnamefont {Mak}}, \ and\ \bibinfo {author}
  {\bibfnamefont {J.}~\bibnamefont {Shan}},\ }\href {\doibase 10.1038/nmat4931}
  {\bibfield  {journal} {\bibinfo  {journal} {Nat. Mater.}\ }\textbf {\bibinfo
  {volume} {16}},\ \bibinfo {pages} {887} (\bibinfo {year} {2017})}\BibitemShut
  {NoStop}%
\bibitem [{\citenamefont {Sodemann}\ and\ \citenamefont
  {Fu}(2015)}]{Sodemann2015}%
  \BibitemOpen
  \bibfield  {author} {\bibinfo {author} {\bibfnamefont {I.}~\bibnamefont
  {Sodemann}}\ and\ \bibinfo {author} {\bibfnamefont {L.}~\bibnamefont {Fu}},\
  }\href {\doibase 10.1103/PhysRevLett.115.216806} {\bibfield  {journal}
  {\bibinfo  {journal} {Phys. Rev. Lett.}\ }\textbf {\bibinfo {volume} {115}},\
  \bibinfo {pages} {216806} (\bibinfo {year} {2015})}\BibitemShut {NoStop}%
\bibitem [{\citenamefont {Yuan}\ \emph {et~al.}(2014)\citenamefont {Yuan},
  \citenamefont {Wang}, \citenamefont {Lian}, \citenamefont {Zhang},
  \citenamefont {Fang}, \citenamefont {Shen}, \citenamefont {Xu}, \citenamefont
  {Xu}, \citenamefont {Zhang}, \citenamefont {Hwang},\ and\ \citenamefont
  {Cui}}]{Yuan2014}%
  \BibitemOpen
  \bibfield  {author} {\bibinfo {author} {\bibfnamefont {H.}~\bibnamefont
  {Yuan}}, \bibinfo {author} {\bibfnamefont {X.}~\bibnamefont {Wang}}, \bibinfo
  {author} {\bibfnamefont {B.}~\bibnamefont {Lian}}, \bibinfo {author}
  {\bibfnamefont {H.}~\bibnamefont {Zhang}}, \bibinfo {author} {\bibfnamefont
  {X.}~\bibnamefont {Fang}}, \bibinfo {author} {\bibfnamefont {B.}~\bibnamefont
  {Shen}}, \bibinfo {author} {\bibfnamefont {G.}~\bibnamefont {Xu}}, \bibinfo
  {author} {\bibfnamefont {Y.}~\bibnamefont {Xu}}, \bibinfo {author}
  {\bibfnamefont {S.-C.}\ \bibnamefont {Zhang}}, \bibinfo {author}
  {\bibfnamefont {H.~Y.}\ \bibnamefont {Hwang}}, \ and\ \bibinfo {author}
  {\bibfnamefont {Y.}~\bibnamefont {Cui}},\ }\href {\doibase
  10.1038/nnano.2014.183} {\bibfield  {journal} {\bibinfo  {journal} {Nat.
  Nanotechnol.}\ }\textbf {\bibinfo {volume} {9}},\ \bibinfo {pages} {851}
  (\bibinfo {year} {2014})}\BibitemShut {NoStop}%
\bibitem [{\citenamefont {Eginligil}\ \emph {et~al.}(2015)\citenamefont
  {Eginligil}, \citenamefont {Cao}, \citenamefont {Wang}, \citenamefont {Shen},
  \citenamefont {Cong}, \citenamefont {Shang}, \citenamefont {Soci},\ and\
  \citenamefont {Yu}}]{Eginligil2015}%
  \BibitemOpen
  \bibfield  {author} {\bibinfo {author} {\bibfnamefont {M.}~\bibnamefont
  {Eginligil}}, \bibinfo {author} {\bibfnamefont {B.}~\bibnamefont {Cao}},
  \bibinfo {author} {\bibfnamefont {Z.}~\bibnamefont {Wang}}, \bibinfo {author}
  {\bibfnamefont {X.}~\bibnamefont {Shen}}, \bibinfo {author} {\bibfnamefont
  {C.}~\bibnamefont {Cong}}, \bibinfo {author} {\bibfnamefont {J.}~\bibnamefont
  {Shang}}, \bibinfo {author} {\bibfnamefont {C.}~\bibnamefont {Soci}}, \ and\
  \bibinfo {author} {\bibfnamefont {T.}~\bibnamefont {Yu}},\ }\href {\doibase
  10.1038/ncomms8636} {\bibfield  {journal} {\bibinfo  {journal} {Nat.
  Commun.}\ }\textbf {\bibinfo {volume} {6}},\ \bibinfo {pages} {7636}
  (\bibinfo {year} {2015})}\BibitemShut {NoStop}%
\bibitem [{\citenamefont {Quereda}\ \emph {et~al.}(2018)\citenamefont
  {Quereda}, \citenamefont {Ghiasi}, \citenamefont {You}, \citenamefont
  {van~den Brink}, \citenamefont {van Wees},\ and\ \citenamefont {van~der
  Wal}}]{Quereda2018}%
  \BibitemOpen
  \bibfield  {author} {\bibinfo {author} {\bibfnamefont {J.}~\bibnamefont
  {Quereda}}, \bibinfo {author} {\bibfnamefont {T.~S.}\ \bibnamefont {Ghiasi}},
  \bibinfo {author} {\bibfnamefont {J.-S.}\ \bibnamefont {You}}, \bibinfo
  {author} {\bibfnamefont {J.}~\bibnamefont {van~den Brink}}, \bibinfo {author}
  {\bibfnamefont {B.~J.}\ \bibnamefont {van Wees}}, \ and\ \bibinfo {author}
  {\bibfnamefont {C.~H.}\ \bibnamefont {van~der Wal}},\ }\href {\doibase
  10.1038/s41467-018-05734-z} {\bibfield  {journal} {\bibinfo  {journal} {Nat.
  Commun.}\ }\textbf {\bibinfo {volume} {9}},\ \bibinfo {pages} {3346}
  (\bibinfo {year} {2018})}\BibitemShut {NoStop}%
\bibitem [{\citenamefont {Zhang}\ \emph
  {et~al.}(2018{\natexlab{a}})\citenamefont {Zhang}, \citenamefont {van~den
  Brink}, \citenamefont {Felser},\ and\ \citenamefont {Yan}}]{Zhang2018b}%
  \BibitemOpen
  \bibfield  {author} {\bibinfo {author} {\bibfnamefont {Y.}~\bibnamefont
  {Zhang}}, \bibinfo {author} {\bibfnamefont {J.}~\bibnamefont {van~den
  Brink}}, \bibinfo {author} {\bibfnamefont {C.}~\bibnamefont {Felser}}, \ and\
  \bibinfo {author} {\bibfnamefont {B.}~\bibnamefont {Yan}},\ }\href {\doibase
  10.1088/2053-1583/aad1ae} {\bibfield  {journal} {\bibinfo  {journal} {2D
  Mater.}\ }\textbf {\bibinfo {volume} {5}},\ \bibinfo {pages} {044001}
  (\bibinfo {year} {2018}{\natexlab{a}})}\BibitemShut {NoStop}%
\bibitem [{\citenamefont {Xu}\ \emph {et~al.}(2018)\citenamefont {Xu},
  \citenamefont {Ma}, \citenamefont {Shen}, \citenamefont {Fatemi},
  \citenamefont {Wu}, \citenamefont {Chang}, \citenamefont {Chang},
  \citenamefont {Valdivia}, \citenamefont {Chan}, \citenamefont {Gibson},
  \citenamefont {Zhou}, \citenamefont {Liu}, \citenamefont {Watanabe},
  \citenamefont {Taniguchi}, \citenamefont {Lin}, \citenamefont {Cava},
  \citenamefont {Fu}, \citenamefont {Gedik},\ and\ \citenamefont
  {Jarillo-Herrero}}]{Xu2018}%
  \BibitemOpen
  \bibfield  {author} {\bibinfo {author} {\bibfnamefont {S.-Y.}\ \bibnamefont
  {Xu}}, \bibinfo {author} {\bibfnamefont {Q.}~\bibnamefont {Ma}}, \bibinfo
  {author} {\bibfnamefont {H.}~\bibnamefont {Shen}}, \bibinfo {author}
  {\bibfnamefont {V.}~\bibnamefont {Fatemi}}, \bibinfo {author} {\bibfnamefont
  {S.}~\bibnamefont {Wu}}, \bibinfo {author} {\bibfnamefont {T.-R.}\
  \bibnamefont {Chang}}, \bibinfo {author} {\bibfnamefont {G.}~\bibnamefont
  {Chang}}, \bibinfo {author} {\bibfnamefont {A.~M.~M.}\ \bibnamefont
  {Valdivia}}, \bibinfo {author} {\bibfnamefont {C.-K.}\ \bibnamefont {Chan}},
  \bibinfo {author} {\bibfnamefont {Q.~D.}\ \bibnamefont {Gibson}}, \bibinfo
  {author} {\bibfnamefont {J.}~\bibnamefont {Zhou}}, \bibinfo {author}
  {\bibfnamefont {Z.}~\bibnamefont {Liu}}, \bibinfo {author} {\bibfnamefont
  {K.}~\bibnamefont {Watanabe}}, \bibinfo {author} {\bibfnamefont
  {T.}~\bibnamefont {Taniguchi}}, \bibinfo {author} {\bibfnamefont
  {H.}~\bibnamefont {Lin}}, \bibinfo {author} {\bibfnamefont {R.~J.}\
  \bibnamefont {Cava}}, \bibinfo {author} {\bibfnamefont {L.}~\bibnamefont
  {Fu}}, \bibinfo {author} {\bibfnamefont {N.}~\bibnamefont {Gedik}}, \ and\
  \bibinfo {author} {\bibfnamefont {P.}~\bibnamefont {Jarillo-Herrero}},\
  }\href {\doibase 10.1038/s41567-018-0189-6} {\bibfield  {journal} {\bibinfo
  {journal} {Nat. Phys.}\ }\textbf {\bibinfo {volume} {14}},\ \bibinfo {pages}
  {900} (\bibinfo {year} {2018})}\BibitemShut {NoStop}%
\bibitem [{\citenamefont {Radisavljevic}\ \emph {et~al.}(2011)\citenamefont
  {Radisavljevic}, \citenamefont {Radenovic}, \citenamefont {Brivio},
  \citenamefont {Giacometti},\ and\ \citenamefont {Kis}}]{Radisavljevic2011}%
  \BibitemOpen
  \bibfield  {author} {\bibinfo {author} {\bibfnamefont {B.}~\bibnamefont
  {Radisavljevic}}, \bibinfo {author} {\bibfnamefont {A.}~\bibnamefont
  {Radenovic}}, \bibinfo {author} {\bibfnamefont {J.}~\bibnamefont {Brivio}},
  \bibinfo {author} {\bibfnamefont {V.}~\bibnamefont {Giacometti}}, \ and\
  \bibinfo {author} {\bibfnamefont {A.}~\bibnamefont {Kis}},\ }\href {\doibase
  10.1038/nnano.2010.279} {\bibfield  {journal} {\bibinfo  {journal} {Nat.
  Nanotechnol.}\ }\textbf {\bibinfo {volume} {6}},\ \bibinfo {pages} {147}
  (\bibinfo {year} {2011})}\BibitemShut {NoStop}%
\bibitem [{\citenamefont {Lopez-Sanchez}\ \emph {et~al.}(2013)\citenamefont
  {Lopez-Sanchez}, \citenamefont {Lembke}, \citenamefont {Kayci}, \citenamefont
  {Radenovic},\ and\ \citenamefont {Kis}}]{Lopez2013}%
  \BibitemOpen
  \bibfield  {author} {\bibinfo {author} {\bibfnamefont {O.}~\bibnamefont
  {Lopez-Sanchez}}, \bibinfo {author} {\bibfnamefont {D.}~\bibnamefont
  {Lembke}}, \bibinfo {author} {\bibfnamefont {M.}~\bibnamefont {Kayci}},
  \bibinfo {author} {\bibfnamefont {A.}~\bibnamefont {Radenovic}}, \ and\
  \bibinfo {author} {\bibfnamefont {A.}~\bibnamefont {Kis}},\ }\href {\doibase
  10.1038/nnano.2013.100} {\bibfield  {journal} {\bibinfo  {journal} {Nat.
  Nanotechnol.}\ }\textbf {\bibinfo {volume} {8}},\ \bibinfo {pages} {497}
  (\bibinfo {year} {2013})}\BibitemShut {NoStop}%
\bibitem [{\citenamefont {Mak}\ \emph {et~al.}(2014)\citenamefont {Mak},
  \citenamefont {McGill}, \citenamefont {Park},\ and\ \citenamefont
  {McEuen}}]{Mak2014}%
  \BibitemOpen
  \bibfield  {author} {\bibinfo {author} {\bibfnamefont {K.~F.}\ \bibnamefont
  {Mak}}, \bibinfo {author} {\bibfnamefont {K.~L.}\ \bibnamefont {McGill}},
  \bibinfo {author} {\bibfnamefont {J.}~\bibnamefont {Park}}, \ and\ \bibinfo
  {author} {\bibfnamefont {P.~L.}\ \bibnamefont {McEuen}},\ }\href {\doibase
  10.1126/science.1250140} {\bibfield  {journal} {\bibinfo  {journal}
  {Science}\ }\textbf {\bibinfo {volume} {344}},\ \bibinfo {pages} {1489}
  (\bibinfo {year} {2014})}\BibitemShut {NoStop}%
\bibitem [{\citenamefont {Qian}\ \emph {et~al.}(2014)\citenamefont {Qian},
  \citenamefont {Liu}, \citenamefont {Fu},\ and\ \citenamefont
  {Li}}]{Qian2014}%
  \BibitemOpen
  \bibfield  {author} {\bibinfo {author} {\bibfnamefont {X.}~\bibnamefont
  {Qian}}, \bibinfo {author} {\bibfnamefont {J.}~\bibnamefont {Liu}}, \bibinfo
  {author} {\bibfnamefont {L.}~\bibnamefont {Fu}}, \ and\ \bibinfo {author}
  {\bibfnamefont {J.}~\bibnamefont {Li}},\ }\href {\doibase
  10.1126/science.1256815} {\bibfield  {journal} {\bibinfo  {journal}
  {Science}\ }\textbf {\bibinfo {volume} {346}},\ \bibinfo {pages} {1344}
  (\bibinfo {year} {2014})}\BibitemShut {NoStop}%
\bibitem [{\citenamefont {You}\ \emph {et~al.}(2018)\citenamefont {You},
  \citenamefont {Fang}, \citenamefont {Xu}, \citenamefont {Kaxiras},\ and\
  \citenamefont {Low}}]{You2018}%
  \BibitemOpen
  \bibfield  {author} {\bibinfo {author} {\bibfnamefont {J.-S.}\ \bibnamefont
  {You}}, \bibinfo {author} {\bibfnamefont {S.}~\bibnamefont {Fang}}, \bibinfo
  {author} {\bibfnamefont {S.-Y.}\ \bibnamefont {Xu}}, \bibinfo {author}
  {\bibfnamefont {E.}~\bibnamefont {Kaxiras}}, \ and\ \bibinfo {author}
  {\bibfnamefont {T.}~\bibnamefont {Low}},\ }\href {\doibase
  10.1103/PhysRevB.98.121109} {\bibfield  {journal} {\bibinfo  {journal} {Phys.
  Rev. B}\ }\textbf {\bibinfo {volume} {98}},\ \bibinfo {pages} {121109}
  (\bibinfo {year} {2018})}\BibitemShut {NoStop}%
\bibitem [{\citenamefont {Young}\ and\ \citenamefont
  {Rappe}(2012)}]{Young2012}%
  \BibitemOpen
  \bibfield  {author} {\bibinfo {author} {\bibfnamefont {S.~M.}\ \bibnamefont
  {Young}}\ and\ \bibinfo {author} {\bibfnamefont {A.~M.}\ \bibnamefont
  {Rappe}},\ }\href {\doibase 10.1103/PhysRevLett.109.116601} {\bibfield
  {journal} {\bibinfo  {journal} {Phys. Rev. Lett.}\ }\textbf {\bibinfo
  {volume} {109}},\ \bibinfo {pages} {116601} (\bibinfo {year}
  {2012})}\BibitemShut {NoStop}%
\bibitem [{\citenamefont {{Iba\~nez-Azpiroz}}\ \emph
  {et~al.}(2018)\citenamefont {{Iba\~nez-Azpiroz}}, \citenamefont {Tsirkin},\
  and\ \citenamefont {Souza}}]{Ibanez2018}%
  \BibitemOpen
  \bibfield  {author} {\bibinfo {author} {\bibfnamefont {J.}~\bibnamefont
  {{Iba\~nez-Azpiroz}}}, \bibinfo {author} {\bibfnamefont {S.~S.}\ \bibnamefont
  {Tsirkin}}, \ and\ \bibinfo {author} {\bibfnamefont {I.}~\bibnamefont
  {Souza}},\ }\href {\doibase 10.1103/PhysRevB.97.245143} {\bibfield  {journal}
  {\bibinfo  {journal} {Phys. Rev. B}\ }\textbf {\bibinfo {volume} {97}},\
  \bibinfo {pages} {245143} (\bibinfo {year} {2018})}\BibitemShut {NoStop}%
\bibitem [{Sup()}]{Supp2018}%
  \BibitemOpen
  \href@noop {} {\enquote {\bibinfo {title} {{See supplemental material at [URL
  will be inserted by APS], which contains the employed expressions for
  calculating the optical conductivities, computed spectra for monolayer
  WSe$_2$, and details of numerical simulations}},}\ }\BibitemShut {NoStop}%
\bibitem [{\citenamefont {Wu}\ \emph {et~al.}(2015)\citenamefont {Wu},
  \citenamefont {Qu},\ and\ \citenamefont {MacDonald}}]{Wu2015}%
  \BibitemOpen
  \bibfield  {author} {\bibinfo {author} {\bibfnamefont {F.}~\bibnamefont
  {Wu}}, \bibinfo {author} {\bibfnamefont {F.}~\bibnamefont {Qu}}, \ and\
  \bibinfo {author} {\bibfnamefont {A.~H.}\ \bibnamefont {MacDonald}},\ }\href
  {\doibase 10.1103/PhysRevB.91.075310} {\bibfield  {journal} {\bibinfo
  {journal} {Phys. Rev. B}\ }\textbf {\bibinfo {volume} {91}},\ \bibinfo
  {pages} {075310} (\bibinfo {year} {2015})}\BibitemShut {NoStop}%
\bibitem [{\citenamefont {Zhou}\ \emph {et~al.}(2015)\citenamefont {Zhou},
  \citenamefont {Shan}, \citenamefont {Yao},\ and\ \citenamefont
  {Xiao}}]{Zhou2015}%
  \BibitemOpen
  \bibfield  {author} {\bibinfo {author} {\bibfnamefont {J.}~\bibnamefont
  {Zhou}}, \bibinfo {author} {\bibfnamefont {W.-Y.}\ \bibnamefont {Shan}},
  \bibinfo {author} {\bibfnamefont {W.}~\bibnamefont {Yao}}, \ and\ \bibinfo
  {author} {\bibfnamefont {D.}~\bibnamefont {Xiao}},\ }\href {\doibase
  10.1103/PhysRevLett.115.166803} {\bibfield  {journal} {\bibinfo  {journal}
  {Phys. Rev. Lett.}\ }\textbf {\bibinfo {volume} {115}},\ \bibinfo {pages}
  {166803} (\bibinfo {year} {2015})}\BibitemShut {NoStop}%
\bibitem [{\citenamefont {Berkelbach}\ \emph {et~al.}(2015)\citenamefont
  {Berkelbach}, \citenamefont {Hybertsen},\ and\ \citenamefont
  {Reichman}}]{Berkelbach2015}%
  \BibitemOpen
  \bibfield  {author} {\bibinfo {author} {\bibfnamefont {T.~C.}\ \bibnamefont
  {Berkelbach}}, \bibinfo {author} {\bibfnamefont {M.~S.}\ \bibnamefont
  {Hybertsen}}, \ and\ \bibinfo {author} {\bibfnamefont {D.~R.}\ \bibnamefont
  {Reichman}},\ }\href {\doibase 10.1103/PhysRevB.92.085413} {\bibfield
  {journal} {\bibinfo  {journal} {Phys. Rev. B}\ }\textbf {\bibinfo {volume}
  {92}},\ \bibinfo {pages} {085413} (\bibinfo {year} {2015})}\BibitemShut
  {NoStop}%
\bibitem [{\citenamefont {Scharf}\ \emph {et~al.}(2017)\citenamefont {Scharf},
  \citenamefont {Xu}, \citenamefont {Matos-Abiague},\ and\ \citenamefont
  {\ifmmode \check{Z}\else \v{Z}\fi{}uti\ifmmode~\acute{c}\else
  \'{c}\fi{}}}]{Scharf2017}%
  \BibitemOpen
  \bibfield  {author} {\bibinfo {author} {\bibfnamefont {B.}~\bibnamefont
  {Scharf}}, \bibinfo {author} {\bibfnamefont {G.}~\bibnamefont {Xu}}, \bibinfo
  {author} {\bibfnamefont {A.}~\bibnamefont {Matos-Abiague}}, \ and\ \bibinfo
  {author} {\bibfnamefont {I.}~\bibnamefont {\ifmmode \check{Z}\else
  \v{Z}\fi{}uti\ifmmode~\acute{c}\else \'{c}\fi{}}},\ }\href {\doibase
  10.1103/PhysRevLett.119.127403} {\bibfield  {journal} {\bibinfo  {journal}
  {Phys. Rev. Lett.}\ }\textbf {\bibinfo {volume} {119}},\ \bibinfo {pages}
  {127403} (\bibinfo {year} {2017})}\BibitemShut {NoStop}%
\bibitem [{\citenamefont {Korm\'anyos}\ \emph {et~al.}(2013)\citenamefont
  {Korm\'anyos}, \citenamefont {Z\'olyomi}, \citenamefont {Drummond},
  \citenamefont {Rakyta}, \citenamefont {Burkard},\ and\ \citenamefont
  {Fal'ko}}]{Kormanyos2013}%
  \BibitemOpen
  \bibfield  {author} {\bibinfo {author} {\bibfnamefont {A.}~\bibnamefont
  {Korm\'anyos}}, \bibinfo {author} {\bibfnamefont {V.}~\bibnamefont
  {Z\'olyomi}}, \bibinfo {author} {\bibfnamefont {N.~D.}\ \bibnamefont
  {Drummond}}, \bibinfo {author} {\bibfnamefont {P.}~\bibnamefont {Rakyta}},
  \bibinfo {author} {\bibfnamefont {G.}~\bibnamefont {Burkard}}, \ and\
  \bibinfo {author} {\bibfnamefont {V.~I.}\ \bibnamefont {Fal'ko}},\ }\href
  {\doibase 10.1103/PhysRevB.88.045416} {\bibfield  {journal} {\bibinfo
  {journal} {Phys. Rev. B}\ }\textbf {\bibinfo {volume} {88}},\ \bibinfo
  {pages} {045416} (\bibinfo {year} {2013})}\BibitemShut {NoStop}%
\bibitem [{\citenamefont {Wang}\ \emph {et~al.}(2018)\citenamefont {Wang},
  \citenamefont {Chernikov}, \citenamefont {Glazov}, \citenamefont {Heinz},
  \citenamefont {Marie}, \citenamefont {Amand},\ and\ \citenamefont
  {Urbaszek}}]{Wang2018}%
  \BibitemOpen
  \bibfield  {author} {\bibinfo {author} {\bibfnamefont {G.}~\bibnamefont
  {Wang}}, \bibinfo {author} {\bibfnamefont {A.}~\bibnamefont {Chernikov}},
  \bibinfo {author} {\bibfnamefont {M.~M.}\ \bibnamefont {Glazov}}, \bibinfo
  {author} {\bibfnamefont {T.~F.}\ \bibnamefont {Heinz}}, \bibinfo {author}
  {\bibfnamefont {X.}~\bibnamefont {Marie}}, \bibinfo {author} {\bibfnamefont
  {T.}~\bibnamefont {Amand}}, \ and\ \bibinfo {author} {\bibfnamefont
  {B.}~\bibnamefont {Urbaszek}},\ }\href {\doibase
  10.1103/RevModPhys.90.021001} {\bibfield  {journal} {\bibinfo  {journal}
  {Rev. Mod. Phys.}\ }\textbf {\bibinfo {volume} {90}},\ \bibinfo {pages}
  {021001} (\bibinfo {year} {2018})}\BibitemShut {NoStop}%
\bibitem [{\citenamefont {Pedersen}(2015)}]{Pedersen2015}%
  \BibitemOpen
  \bibfield  {author} {\bibinfo {author} {\bibfnamefont {T.~G.}\ \bibnamefont
  {Pedersen}},\ }\href {\doibase 10.1103/PhysRevB.92.235432} {\bibfield
  {journal} {\bibinfo  {journal} {Phys. Rev. B}\ }\textbf {\bibinfo {volume}
  {92}},\ \bibinfo {pages} {235432} (\bibinfo {year} {2015})}\BibitemShut
  {NoStop}%
\bibitem [{\citenamefont {Taghizadeh}\ and\ \citenamefont
  {Pedersen}(2018)}]{Taghizadeh2018}%
  \BibitemOpen
  \bibfield  {author} {\bibinfo {author} {\bibfnamefont {A.}~\bibnamefont
  {Taghizadeh}}\ and\ \bibinfo {author} {\bibfnamefont {T.~G.}\ \bibnamefont
  {Pedersen}},\ }\href {\doibase 10.1103/PhysRevB.97.205432} {\bibfield
  {journal} {\bibinfo  {journal} {Phys. Rev. B}\ }\textbf {\bibinfo {volume}
  {97}},\ \bibinfo {pages} {205432} (\bibinfo {year} {2018})}\BibitemShut
  {NoStop}%
\bibitem [{\citenamefont {Albrecht}\ \emph {et~al.}(1998)\citenamefont
  {Albrecht}, \citenamefont {Reining}, \citenamefont {Del~Sole},\ and\
  \citenamefont {Onida}}]{Albrecht1998}%
  \BibitemOpen
  \bibfield  {author} {\bibinfo {author} {\bibfnamefont {S.}~\bibnamefont
  {Albrecht}}, \bibinfo {author} {\bibfnamefont {L.}~\bibnamefont {Reining}},
  \bibinfo {author} {\bibfnamefont {R.}~\bibnamefont {Del~Sole}}, \ and\
  \bibinfo {author} {\bibfnamefont {G.}~\bibnamefont {Onida}},\ }\href
  {\doibase 10.1103/PhysRevLett.80.4510} {\bibfield  {journal} {\bibinfo
  {journal} {Phys. Rev. Lett.}\ }\textbf {\bibinfo {volume} {80}},\ \bibinfo
  {pages} {4510} (\bibinfo {year} {1998})}\BibitemShut {NoStop}%
\bibitem [{\citenamefont {Rohlfing}\ and\ \citenamefont
  {Louie}(2000)}]{Rohlfing2000}%
  \BibitemOpen
  \bibfield  {author} {\bibinfo {author} {\bibfnamefont {M.}~\bibnamefont
  {Rohlfing}}\ and\ \bibinfo {author} {\bibfnamefont {S.~G.}\ \bibnamefont
  {Louie}},\ }\href {\doibase 10.1103/PhysRevB.62.4927} {\bibfield  {journal}
  {\bibinfo  {journal} {Phys. Rev. B}\ }\textbf {\bibinfo {volume} {62}},\
  \bibinfo {pages} {4927} (\bibinfo {year} {2000})}\BibitemShut {NoStop}%
\bibitem [{\citenamefont {Cudazzo}\ \emph {et~al.}(2011)\citenamefont
  {Cudazzo}, \citenamefont {Tokatly},\ and\ \citenamefont
  {Rubio}}]{Cudazzo2011}%
  \BibitemOpen
  \bibfield  {author} {\bibinfo {author} {\bibfnamefont {P.}~\bibnamefont
  {Cudazzo}}, \bibinfo {author} {\bibfnamefont {I.~V.}\ \bibnamefont
  {Tokatly}}, \ and\ \bibinfo {author} {\bibfnamefont {A.}~\bibnamefont
  {Rubio}},\ }\href {\doibase 10.1103/PhysRevB.84.085406} {\bibfield  {journal}
  {\bibinfo  {journal} {Phys. Rev. B}\ }\textbf {\bibinfo {volume} {84}},\
  \bibinfo {pages} {085406} (\bibinfo {year} {2011})}\BibitemShut {NoStop}%
\bibitem [{\citenamefont {Wang}\ \emph {et~al.}(2015)\citenamefont {Wang},
  \citenamefont {Marie}, \citenamefont {Gerber}, \citenamefont {Amand},
  \citenamefont {Lagarde}, \citenamefont {Bouet}, \citenamefont {Vidal},
  \citenamefont {Balocchi},\ and\ \citenamefont {Urbaszek}}]{Wang2015}%
  \BibitemOpen
  \bibfield  {author} {\bibinfo {author} {\bibfnamefont {G.}~\bibnamefont
  {Wang}}, \bibinfo {author} {\bibfnamefont {X.}~\bibnamefont {Marie}},
  \bibinfo {author} {\bibfnamefont {I.}~\bibnamefont {Gerber}}, \bibinfo
  {author} {\bibfnamefont {T.}~\bibnamefont {Amand}}, \bibinfo {author}
  {\bibfnamefont {D.}~\bibnamefont {Lagarde}}, \bibinfo {author} {\bibfnamefont
  {L.}~\bibnamefont {Bouet}}, \bibinfo {author} {\bibfnamefont
  {M.}~\bibnamefont {Vidal}}, \bibinfo {author} {\bibfnamefont
  {A.}~\bibnamefont {Balocchi}}, \ and\ \bibinfo {author} {\bibfnamefont
  {B.}~\bibnamefont {Urbaszek}},\ }\href {\doibase
  10.1103/PhysRevLett.114.097403} {\bibfield  {journal} {\bibinfo  {journal}
  {Phys. Rev. Lett.}\ }\textbf {\bibinfo {volume} {114}},\ \bibinfo {pages}
  {097403} (\bibinfo {year} {2015})}\BibitemShut {NoStop}%
\bibitem [{\citenamefont {Srivastava}\ and\ \citenamefont
  {Imamo{\u{g}}lu}(2015)}]{Srivastava2015}%
  \BibitemOpen
  \bibfield  {author} {\bibinfo {author} {\bibfnamefont {A.}~\bibnamefont
  {Srivastava}}\ and\ \bibinfo {author} {\bibfnamefont {A.}~\bibnamefont
  {Imamo{\u{g}}lu}},\ }\href {\doibase 10.1103/PhysRevLett.115.166802}
  {\bibfield  {journal} {\bibinfo  {journal} {Phys. Rev. Lett.}\ }\textbf
  {\bibinfo {volume} {115}},\ \bibinfo {pages} {166802} (\bibinfo {year}
  {2015})}\BibitemShut {NoStop}%
\bibitem [{\citenamefont {Taghizadeh}\ \emph {et~al.}(2017)\citenamefont
  {Taghizadeh}, \citenamefont {Hipolito},\ and\ \citenamefont
  {Pedersen}}]{Taghizadeh2017}%
  \BibitemOpen
  \bibfield  {author} {\bibinfo {author} {\bibfnamefont {A.}~\bibnamefont
  {Taghizadeh}}, \bibinfo {author} {\bibfnamefont {F.}~\bibnamefont
  {Hipolito}}, \ and\ \bibinfo {author} {\bibfnamefont {T.~G.}\ \bibnamefont
  {Pedersen}},\ }\href {\doibase 10.1103/PhysRevB.96.195413} {\bibfield
  {journal} {\bibinfo  {journal} {Phys. Rev. B}\ }\textbf {\bibinfo {volume}
  {96}},\ \bibinfo {pages} {195413} (\bibinfo {year} {2017})}\BibitemShut
  {NoStop}%
\bibitem [{\citenamefont {Hipolito}\ \emph {et~al.}(2018)\citenamefont
  {Hipolito}, \citenamefont {Taghizadeh},\ and\ \citenamefont
  {Pedersen}}]{Hipolito2018}%
  \BibitemOpen
  \bibfield  {author} {\bibinfo {author} {\bibfnamefont {F.}~\bibnamefont
  {Hipolito}}, \bibinfo {author} {\bibfnamefont {A.}~\bibnamefont
  {Taghizadeh}}, \ and\ \bibinfo {author} {\bibfnamefont {T.~G.}\ \bibnamefont
  {Pedersen}},\ }\href {\doibase 10.1103/PhysRevB.98.205420} {\bibfield
  {journal} {\bibinfo  {journal} {Phys. Rev. B}\ }\textbf {\bibinfo {volume}
  {98}},\ \bibinfo {pages} {205420} (\bibinfo {year} {2018})}\BibitemShut
  {NoStop}%
\bibitem [{\citenamefont {Passos}\ \emph {et~al.}(2018)\citenamefont {Passos},
  \citenamefont {Ventura}, \citenamefont {Lopes}, \citenamefont {dos Santos},\
  and\ \citenamefont {Peres}}]{Passos2018}%
  \BibitemOpen
  \bibfield  {author} {\bibinfo {author} {\bibfnamefont {D.~J.}\ \bibnamefont
  {Passos}}, \bibinfo {author} {\bibfnamefont {G.~B.}\ \bibnamefont {Ventura}},
  \bibinfo {author} {\bibfnamefont {J.~M. V.~P.}\ \bibnamefont {Lopes}},
  \bibinfo {author} {\bibfnamefont {J.~M. B.~L.}\ \bibnamefont {dos Santos}}, \
  and\ \bibinfo {author} {\bibfnamefont {N.~M.~R.}\ \bibnamefont {Peres}},\
  }\href {\doibase 10.1103/PhysRevB.97.235446} {\bibfield  {journal} {\bibinfo
  {journal} {Phys. Rev. B}\ }\textbf {\bibinfo {volume} {97}},\ \bibinfo
  {pages} {235446} (\bibinfo {year} {2018})}\BibitemShut {NoStop}%
\bibitem [{\citenamefont {Mermin}(1970)}]{Mermin1970}%
  \BibitemOpen
  \bibfield  {author} {\bibinfo {author} {\bibfnamefont {N.~D.}\ \bibnamefont
  {Mermin}},\ }\href {\doibase 10.1103/PhysRevB.1.2362} {\bibfield  {journal}
  {\bibinfo  {journal} {Phys. Rev. B}\ }\textbf {\bibinfo {volume} {1}},\
  \bibinfo {pages} {2362} (\bibinfo {year} {1970})}\BibitemShut {NoStop}%
\bibitem [{\citenamefont {Cheng}\ \emph {et~al.}(2015)\citenamefont {Cheng},
  \citenamefont {Vermeulen},\ and\ \citenamefont {Sipe}}]{Cheng2015}%
  \BibitemOpen
  \bibfield  {author} {\bibinfo {author} {\bibfnamefont {J.~L.}\ \bibnamefont
  {Cheng}}, \bibinfo {author} {\bibfnamefont {N.}~\bibnamefont {Vermeulen}}, \
  and\ \bibinfo {author} {\bibfnamefont {J.~E.}\ \bibnamefont {Sipe}},\ }\href
  {\doibase 10.1103/PhysRevB.91.235320} {\bibfield  {journal} {\bibinfo
  {journal} {Phys. Rev. B}\ }\textbf {\bibinfo {volume} {91}},\ \bibinfo
  {pages} {235320} (\bibinfo {year} {2015})}\BibitemShut {NoStop}%
\bibitem [{\citenamefont {Hipolito}\ \emph {et~al.}(2016)\citenamefont
  {Hipolito}, \citenamefont {Pedersen},\ and\ \citenamefont
  {Pereira}}]{Hipolito2016}%
  \BibitemOpen
  \bibfield  {author} {\bibinfo {author} {\bibfnamefont {F.}~\bibnamefont
  {Hipolito}}, \bibinfo {author} {\bibfnamefont {T.~G.}\ \bibnamefont
  {Pedersen}}, \ and\ \bibinfo {author} {\bibfnamefont {V.~M.}\ \bibnamefont
  {Pereira}},\ }\href {\doibase 10.1103/PhysRevB.94.045434} {\bibfield
  {journal} {\bibinfo  {journal} {Phys. Rev. B}\ }\textbf {\bibinfo {volume}
  {94}},\ \bibinfo {pages} {045434} (\bibinfo {year} {2016})}\BibitemShut
  {NoStop}%
\bibitem [{\citenamefont {Cadiz}\ \emph {et~al.}(2017)\citenamefont {Cadiz},
  \citenamefont {Courtade}, \citenamefont {Robert}, \citenamefont {Wang},
  \citenamefont {Shen}, \citenamefont {Cai}, \citenamefont {Taniguchi},
  \citenamefont {Watanabe}, \citenamefont {Carrere}, \citenamefont {Lagarde},
  \citenamefont {Manca}, \citenamefont {Amand}, \citenamefont {Renucci},
  \citenamefont {Tongay}, \citenamefont {Marie},\ and\ \citenamefont
  {Urbaszek}}]{Cadiz2017}%
  \BibitemOpen
  \bibfield  {author} {\bibinfo {author} {\bibfnamefont {F.}~\bibnamefont
  {Cadiz}}, \bibinfo {author} {\bibfnamefont {E.}~\bibnamefont {Courtade}},
  \bibinfo {author} {\bibfnamefont {C.}~\bibnamefont {Robert}}, \bibinfo
  {author} {\bibfnamefont {G.}~\bibnamefont {Wang}}, \bibinfo {author}
  {\bibfnamefont {Y.}~\bibnamefont {Shen}}, \bibinfo {author} {\bibfnamefont
  {H.}~\bibnamefont {Cai}}, \bibinfo {author} {\bibfnamefont {T.}~\bibnamefont
  {Taniguchi}}, \bibinfo {author} {\bibfnamefont {K.}~\bibnamefont {Watanabe}},
  \bibinfo {author} {\bibfnamefont {H.}~\bibnamefont {Carrere}}, \bibinfo
  {author} {\bibfnamefont {D.}~\bibnamefont {Lagarde}}, \bibinfo {author}
  {\bibfnamefont {M.}~\bibnamefont {Manca}}, \bibinfo {author} {\bibfnamefont
  {T.}~\bibnamefont {Amand}}, \bibinfo {author} {\bibfnamefont
  {P.}~\bibnamefont {Renucci}}, \bibinfo {author} {\bibfnamefont
  {S.}~\bibnamefont {Tongay}}, \bibinfo {author} {\bibfnamefont
  {X.}~\bibnamefont {Marie}}, \ and\ \bibinfo {author} {\bibfnamefont
  {B.}~\bibnamefont {Urbaszek}},\ }\href {\doibase 10.1103/PhysRevX.7.021026}
  {\bibfield  {journal} {\bibinfo  {journal} {Phys. Rev. X}\ }\textbf {\bibinfo
  {volume} {7}},\ \bibinfo {pages} {021026} (\bibinfo {year}
  {2017})}\BibitemShut {NoStop}%
\bibitem [{\citenamefont {Selig}\ \emph {et~al.}(2016)\citenamefont {Selig},
  \citenamefont {Bergh{\"a}user}, \citenamefont {Raja}, \citenamefont {Nagler},
  \citenamefont {Sch{\"u}ller}, \citenamefont {Heinz}, \citenamefont {Korn},
  \citenamefont {Chernikov}, \citenamefont {Malic},\ and\ \citenamefont
  {Knorr}}]{Selig2016}%
  \BibitemOpen
  \bibfield  {author} {\bibinfo {author} {\bibfnamefont {M.}~\bibnamefont
  {Selig}}, \bibinfo {author} {\bibfnamefont {G.}~\bibnamefont
  {Bergh{\"a}user}}, \bibinfo {author} {\bibfnamefont {A.}~\bibnamefont
  {Raja}}, \bibinfo {author} {\bibfnamefont {P.}~\bibnamefont {Nagler}},
  \bibinfo {author} {\bibfnamefont {C.}~\bibnamefont {Sch{\"u}ller}}, \bibinfo
  {author} {\bibfnamefont {T.~F.}\ \bibnamefont {Heinz}}, \bibinfo {author}
  {\bibfnamefont {T.}~\bibnamefont {Korn}}, \bibinfo {author} {\bibfnamefont
  {A.}~\bibnamefont {Chernikov}}, \bibinfo {author} {\bibfnamefont
  {E.}~\bibnamefont {Malic}}, \ and\ \bibinfo {author} {\bibfnamefont
  {A.}~\bibnamefont {Knorr}},\ }\href {\doibase 10.1038/ncomms13279} {\bibfield
   {journal} {\bibinfo  {journal} {Nat. Commun.}\ }\textbf {\bibinfo {volume}
  {7}},\ \bibinfo {pages} {13279} (\bibinfo {year} {2016})}\BibitemShut
  {NoStop}%
\bibitem [{\citenamefont {Christiansen}\ \emph {et~al.}(2017)\citenamefont
  {Christiansen}, \citenamefont {Selig}, \citenamefont {Bergh\"auser},
  \citenamefont {Schmidt}, \citenamefont {Niehues}, \citenamefont {Schneider},
  \citenamefont {Arora}, \citenamefont {de~Vasconcellos}, \citenamefont
  {Bratschitsch}, \citenamefont {Malic},\ and\ \citenamefont
  {Knorr}}]{Christiansen2017}%
  \BibitemOpen
  \bibfield  {author} {\bibinfo {author} {\bibfnamefont {D.}~\bibnamefont
  {Christiansen}}, \bibinfo {author} {\bibfnamefont {M.}~\bibnamefont {Selig}},
  \bibinfo {author} {\bibfnamefont {G.}~\bibnamefont {Bergh\"auser}}, \bibinfo
  {author} {\bibfnamefont {R.}~\bibnamefont {Schmidt}}, \bibinfo {author}
  {\bibfnamefont {I.}~\bibnamefont {Niehues}}, \bibinfo {author} {\bibfnamefont
  {R.}~\bibnamefont {Schneider}}, \bibinfo {author} {\bibfnamefont
  {A.}~\bibnamefont {Arora}}, \bibinfo {author} {\bibfnamefont {S.~M.}\
  \bibnamefont {de~Vasconcellos}}, \bibinfo {author} {\bibfnamefont
  {R.}~\bibnamefont {Bratschitsch}}, \bibinfo {author} {\bibfnamefont
  {E.}~\bibnamefont {Malic}}, \ and\ \bibinfo {author} {\bibfnamefont
  {A.}~\bibnamefont {Knorr}},\ }\href {\doibase 10.1103/PhysRevLett.119.187402}
  {\bibfield  {journal} {\bibinfo  {journal} {Phys. Rev. Lett.}\ }\textbf
  {\bibinfo {volume} {119}},\ \bibinfo {pages} {187402} (\bibinfo {year}
  {2017})}\BibitemShut {NoStop}%
\bibitem [{\citenamefont {Scuri}\ \emph {et~al.}(2018)\citenamefont {Scuri},
  \citenamefont {Zhou}, \citenamefont {High}, \citenamefont {Wild},
  \citenamefont {Shu}, \citenamefont {De~Greve}, \citenamefont {Jauregui},
  \citenamefont {Taniguchi}, \citenamefont {Watanabe}, \citenamefont {Kim},
  \citenamefont {Lukin},\ and\ \citenamefont {Park}}]{Scuri2018}%
  \BibitemOpen
  \bibfield  {author} {\bibinfo {author} {\bibfnamefont {G.}~\bibnamefont
  {Scuri}}, \bibinfo {author} {\bibfnamefont {Y.}~\bibnamefont {Zhou}},
  \bibinfo {author} {\bibfnamefont {A.~A.}\ \bibnamefont {High}}, \bibinfo
  {author} {\bibfnamefont {D.~S.}\ \bibnamefont {Wild}}, \bibinfo {author}
  {\bibfnamefont {C.}~\bibnamefont {Shu}}, \bibinfo {author} {\bibfnamefont
  {K.}~\bibnamefont {De~Greve}}, \bibinfo {author} {\bibfnamefont {L.~A.}\
  \bibnamefont {Jauregui}}, \bibinfo {author} {\bibfnamefont {T.}~\bibnamefont
  {Taniguchi}}, \bibinfo {author} {\bibfnamefont {K.}~\bibnamefont {Watanabe}},
  \bibinfo {author} {\bibfnamefont {P.}~\bibnamefont {Kim}}, \bibinfo {author}
  {\bibfnamefont {M.~D.}\ \bibnamefont {Lukin}}, \ and\ \bibinfo {author}
  {\bibfnamefont {H.}~\bibnamefont {Park}},\ }\href {\doibase
  10.1103/PhysRevLett.120.037402} {\bibfield  {journal} {\bibinfo  {journal}
  {Phys. Rev. Lett.}\ }\textbf {\bibinfo {volume} {120}},\ \bibinfo {pages}
  {037402} (\bibinfo {year} {2018})}\BibitemShut {NoStop}%
\bibitem [{\citenamefont {Shen}\ \emph {et~al.}(2013)\citenamefont {Shen},
  \citenamefont {Hsu}, \citenamefont {Li},\ and\ \citenamefont
  {Liu}}]{Shen2013}%
  \BibitemOpen
  \bibfield  {author} {\bibinfo {author} {\bibfnamefont {C.-C.}\ \bibnamefont
  {Shen}}, \bibinfo {author} {\bibfnamefont {Y.-T.}\ \bibnamefont {Hsu}},
  \bibinfo {author} {\bibfnamefont {L.-J.}\ \bibnamefont {Li}}, \ and\ \bibinfo
  {author} {\bibfnamefont {H.-L.}\ \bibnamefont {Liu}},\ }\href {\doibase
  10.7567/apex.6.125801} {\bibfield  {journal} {\bibinfo  {journal} {Appl.
  Phys. Express}\ }\textbf {\bibinfo {volume} {6}},\ \bibinfo {pages} {125801}
  (\bibinfo {year} {2013})}\BibitemShut {NoStop}%
\bibitem [{\citenamefont {Gupta}\ \emph {et~al.}(2018)\citenamefont {Gupta},
  \citenamefont {Shirodkar}, \citenamefont {Kutana},\ and\ \citenamefont
  {Yakobson}}]{Gupta2018}%
  \BibitemOpen
  \bibfield  {author} {\bibinfo {author} {\bibfnamefont {S.}~\bibnamefont
  {Gupta}}, \bibinfo {author} {\bibfnamefont {S.~N.}\ \bibnamefont
  {Shirodkar}}, \bibinfo {author} {\bibfnamefont {A.}~\bibnamefont {Kutana}}, \
  and\ \bibinfo {author} {\bibfnamefont {B.~I.}\ \bibnamefont {Yakobson}},\
  }\href {\doibase 10.1021/acsnano.8b03754} {\bibfield  {journal} {\bibinfo
  {journal} {{ACS} Nano}\ }\textbf {\bibinfo {volume} {12}},\ \bibinfo {pages}
  {10880} (\bibinfo {year} {2018})}\BibitemShut {NoStop}%
\bibitem [{\citenamefont {Ye}\ \emph {et~al.}(2014)\citenamefont {Ye},
  \citenamefont {Cao}, \citenamefont {O'Brien}, \citenamefont {Zhu},
  \citenamefont {Yin}, \citenamefont {Wang}, \citenamefont {Louie},\ and\
  \citenamefont {Zhang}}]{Ye2014}%
  \BibitemOpen
  \bibfield  {author} {\bibinfo {author} {\bibfnamefont {Z.}~\bibnamefont
  {Ye}}, \bibinfo {author} {\bibfnamefont {T.}~\bibnamefont {Cao}}, \bibinfo
  {author} {\bibfnamefont {K.}~\bibnamefont {O'Brien}}, \bibinfo {author}
  {\bibfnamefont {H.}~\bibnamefont {Zhu}}, \bibinfo {author} {\bibfnamefont
  {X.}~\bibnamefont {Yin}}, \bibinfo {author} {\bibfnamefont {Y.}~\bibnamefont
  {Wang}}, \bibinfo {author} {\bibfnamefont {S.~G.}\ \bibnamefont {Louie}}, \
  and\ \bibinfo {author} {\bibfnamefont {X.}~\bibnamefont {Zhang}},\ }\href
  {\doibase 10.1038/nature13734} {\bibfield  {journal} {\bibinfo  {journal}
  {Nature}\ }\textbf {\bibinfo {volume} {513}},\ \bibinfo {pages} {214}
  (\bibinfo {year} {2014})}\BibitemShut {NoStop}%
\bibitem [{\citenamefont {Cao}\ \emph {et~al.}(2018)\citenamefont {Cao},
  \citenamefont {Wu},\ and\ \citenamefont {Louie}}]{Cao2018}%
  \BibitemOpen
  \bibfield  {author} {\bibinfo {author} {\bibfnamefont {T.}~\bibnamefont
  {Cao}}, \bibinfo {author} {\bibfnamefont {M.}~\bibnamefont {Wu}}, \ and\
  \bibinfo {author} {\bibfnamefont {S.~G.}\ \bibnamefont {Louie}},\ }\href
  {\doibase 10.1103/PhysRevLett.120.087402} {\bibfield  {journal} {\bibinfo
  {journal} {Phys. Rev. Lett.}\ }\textbf {\bibinfo {volume} {120}},\ \bibinfo
  {pages} {087402} (\bibinfo {year} {2018})}\BibitemShut {NoStop}%
\bibitem [{\citenamefont {Zhang}\ \emph
  {et~al.}(2018{\natexlab{b}})\citenamefont {Zhang}, \citenamefont {Shan},\
  and\ \citenamefont {Xiao}}]{Zhang2018}%
  \BibitemOpen
  \bibfield  {author} {\bibinfo {author} {\bibfnamefont {X.}~\bibnamefont
  {Zhang}}, \bibinfo {author} {\bibfnamefont {W.-Y.}\ \bibnamefont {Shan}}, \
  and\ \bibinfo {author} {\bibfnamefont {D.}~\bibnamefont {Xiao}},\ }\href
  {\doibase 10.1103/PhysRevLett.120.077401} {\bibfield  {journal} {\bibinfo
  {journal} {Phys. Rev. Lett.}\ }\textbf {\bibinfo {volume} {120}},\ \bibinfo
  {pages} {077401} (\bibinfo {year} {2018}{\natexlab{b}})}\BibitemShut
  {NoStop}%
\bibitem [{\citenamefont {Seyler}\ \emph {et~al.}(2015)\citenamefont {Seyler},
  \citenamefont {Schaibley}, \citenamefont {Gong}, \citenamefont {Rivera},
  \citenamefont {Jones}, \citenamefont {Wu}, \citenamefont {Yan}, \citenamefont
  {Mandrus}, \citenamefont {Yao},\ and\ \citenamefont {Xu}}]{Seyler2015}%
  \BibitemOpen
  \bibfield  {author} {\bibinfo {author} {\bibfnamefont {K.~L.}\ \bibnamefont
  {Seyler}}, \bibinfo {author} {\bibfnamefont {J.~R.}\ \bibnamefont
  {Schaibley}}, \bibinfo {author} {\bibfnamefont {P.}~\bibnamefont {Gong}},
  \bibinfo {author} {\bibfnamefont {P.}~\bibnamefont {Rivera}}, \bibinfo
  {author} {\bibfnamefont {A.~M.}\ \bibnamefont {Jones}}, \bibinfo {author}
  {\bibfnamefont {S.}~\bibnamefont {Wu}}, \bibinfo {author} {\bibfnamefont
  {J.}~\bibnamefont {Yan}}, \bibinfo {author} {\bibfnamefont {D.~G.}\
  \bibnamefont {Mandrus}}, \bibinfo {author} {\bibfnamefont {W.}~\bibnamefont
  {Yao}}, \ and\ \bibinfo {author} {\bibfnamefont {X.}~\bibnamefont {Xu}},\
  }\href {\doibase 10.1038/nnano.2015.73} {\bibfield  {journal} {\bibinfo
  {journal} {Nat. Nanotechnol.}\ }\textbf {\bibinfo {volume} {10}},\ \bibinfo
  {pages} {407} (\bibinfo {year} {2015})}\BibitemShut {NoStop}%
\bibitem [{\citenamefont {Zhao}\ \emph {et~al.}(2006)\citenamefont {Zhao},
  \citenamefont {Loren}, \citenamefont {van Driel},\ and\ \citenamefont
  {Smirl}}]{Zhao2006}%
  \BibitemOpen
  \bibfield  {author} {\bibinfo {author} {\bibfnamefont {H.}~\bibnamefont
  {Zhao}}, \bibinfo {author} {\bibfnamefont {E.~J.}\ \bibnamefont {Loren}},
  \bibinfo {author} {\bibfnamefont {H.~M.}\ \bibnamefont {van Driel}}, \ and\
  \bibinfo {author} {\bibfnamefont {A.~L.}\ \bibnamefont {Smirl}},\ }\href
  {\doibase 10.1103/PhysRevLett.96.246601} {\bibfield  {journal} {\bibinfo
  {journal} {Phys. Rev. Lett.}\ }\textbf {\bibinfo {volume} {96}},\ \bibinfo
  {pages} {246601} (\bibinfo {year} {2006})}\BibitemShut {NoStop}%
\bibitem [{\citenamefont {Wang}\ \emph {et~al.}(2010)\citenamefont {Wang},
  \citenamefont {Zhu},\ and\ \citenamefont {Liu}}]{Wang2010}%
  \BibitemOpen
  \bibfield  {author} {\bibinfo {author} {\bibfnamefont {J.}~\bibnamefont
  {Wang}}, \bibinfo {author} {\bibfnamefont {B.-F.}\ \bibnamefont {Zhu}}, \
  and\ \bibinfo {author} {\bibfnamefont {R.-B.}\ \bibnamefont {Liu}},\ }\href
  {\doibase 10.1103/PhysRevLett.104.256601} {\bibfield  {journal} {\bibinfo
  {journal} {Phys. Rev. Lett.}\ }\textbf {\bibinfo {volume} {104}},\ \bibinfo
  {pages} {256601} (\bibinfo {year} {2010})}\BibitemShut {NoStop}%
\bibitem [{\citenamefont {Werake}\ and\ \citenamefont
  {Zhao}(2010)}]{Werake2010}%
  \BibitemOpen
  \bibfield  {author} {\bibinfo {author} {\bibfnamefont {L.~K.}\ \bibnamefont
  {Werake}}\ and\ \bibinfo {author} {\bibfnamefont {H.}~\bibnamefont {Zhao}},\
  }\href {\doibase 10.1038/nphys1742} {\bibfield  {journal} {\bibinfo
  {journal} {Nat. Phys.}\ }\textbf {\bibinfo {volume} {6}},\ \bibinfo {pages}
  {875} (\bibinfo {year} {2010})}\BibitemShut {NoStop}%
\bibitem [{\citenamefont {Zeng}\ \emph {et~al.}(2012)\citenamefont {Zeng},
  \citenamefont {Dai}, \citenamefont {Yao}, \citenamefont {Xiao},\ and\
  \citenamefont {Cui}}]{Zeng2012}%
  \BibitemOpen
  \bibfield  {author} {\bibinfo {author} {\bibfnamefont {H.}~\bibnamefont
  {Zeng}}, \bibinfo {author} {\bibfnamefont {J.}~\bibnamefont {Dai}}, \bibinfo
  {author} {\bibfnamefont {W.}~\bibnamefont {Yao}}, \bibinfo {author}
  {\bibfnamefont {D.}~\bibnamefont {Xiao}}, \ and\ \bibinfo {author}
  {\bibfnamefont {X.}~\bibnamefont {Cui}},\ }\href {\doibase
  10.1038/nnano.2012.95} {\bibfield  {journal} {\bibinfo  {journal} {Nat.
  Nanotechnol.}\ }\textbf {\bibinfo {volume} {7}},\ \bibinfo {pages} {490}
  (\bibinfo {year} {2012})}\BibitemShut {NoStop}%
\bibitem [{\citenamefont {Mak}\ \emph {et~al.}(2012)\citenamefont {Mak},
  \citenamefont {He}, \citenamefont {Shan},\ and\ \citenamefont
  {Heinz}}]{Mak2012}%
  \BibitemOpen
  \bibfield  {author} {\bibinfo {author} {\bibfnamefont {K.~F.}\ \bibnamefont
  {Mak}}, \bibinfo {author} {\bibfnamefont {K.}~\bibnamefont {He}}, \bibinfo
  {author} {\bibfnamefont {J.}~\bibnamefont {Shan}}, \ and\ \bibinfo {author}
  {\bibfnamefont {T.~F.}\ \bibnamefont {Heinz}},\ }\href {\doibase
  10.1038/nnano.2012.96} {\bibfield  {journal} {\bibinfo  {journal} {Nat.
  Nanotechnol.}\ }\textbf {\bibinfo {volume} {7}},\ \bibinfo {pages} {494}
  (\bibinfo {year} {2012})}\BibitemShut {NoStop}%
\bibitem [{\citenamefont {Cao}\ \emph {et~al.}(2012)\citenamefont {Cao},
  \citenamefont {Wang}, \citenamefont {Han}, \citenamefont {Ye}, \citenamefont
  {Zhu}, \citenamefont {Shi}, \citenamefont {Niu}, \citenamefont {Tan},
  \citenamefont {Wang}, \citenamefont {Liu},\ and\ \citenamefont
  {Feng}}]{Cao2012}%
  \BibitemOpen
  \bibfield  {author} {\bibinfo {author} {\bibfnamefont {T.}~\bibnamefont
  {Cao}}, \bibinfo {author} {\bibfnamefont {G.}~\bibnamefont {Wang}}, \bibinfo
  {author} {\bibfnamefont {W.}~\bibnamefont {Han}}, \bibinfo {author}
  {\bibfnamefont {H.}~\bibnamefont {Ye}}, \bibinfo {author} {\bibfnamefont
  {C.}~\bibnamefont {Zhu}}, \bibinfo {author} {\bibfnamefont {J.}~\bibnamefont
  {Shi}}, \bibinfo {author} {\bibfnamefont {Q.}~\bibnamefont {Niu}}, \bibinfo
  {author} {\bibfnamefont {P.}~\bibnamefont {Tan}}, \bibinfo {author}
  {\bibfnamefont {E.}~\bibnamefont {Wang}}, \bibinfo {author} {\bibfnamefont
  {B.}~\bibnamefont {Liu}}, \ and\ \bibinfo {author} {\bibfnamefont
  {J.}~\bibnamefont {Feng}},\ }\href {\doibase 10.1038/ncomms1882} {\bibfield
  {journal} {\bibinfo  {journal} {Nat. Commun.}\ }\textbf {\bibinfo {volume}
  {3}},\ \bibinfo {pages} {887} (\bibinfo {year} {2012})}\BibitemShut {NoStop}%
\bibitem [{\citenamefont {Xiao}\ \emph {et~al.}(2015)\citenamefont {Xiao},
  \citenamefont {Ye}, \citenamefont {Wang}, \citenamefont {Zhu}, \citenamefont
  {Wang},\ and\ \citenamefont {Zhang}}]{Xiao2015}%
  \BibitemOpen
  \bibfield  {author} {\bibinfo {author} {\bibfnamefont {J.}~\bibnamefont
  {Xiao}}, \bibinfo {author} {\bibfnamefont {Z.}~\bibnamefont {Ye}}, \bibinfo
  {author} {\bibfnamefont {Y.}~\bibnamefont {Wang}}, \bibinfo {author}
  {\bibfnamefont {H.}~\bibnamefont {Zhu}}, \bibinfo {author} {\bibfnamefont
  {Y.}~\bibnamefont {Wang}}, \ and\ \bibinfo {author} {\bibfnamefont
  {X.}~\bibnamefont {Zhang}},\ }\href {\doibase 10.1038/lsa.2015.139}
  {\bibfield  {journal} {\bibinfo  {journal} {Light Sci. Appl.}\ }\textbf
  {\bibinfo {volume} {4}},\ \bibinfo {pages} {e366} (\bibinfo {year}
  {2015})}\BibitemShut {NoStop}%
\end{thebibliography}%


\begin{thebibliography}{19}%
\makeatletter
\providecommand \@ifxundefined [1]{%
 \@ifx{#1\undefined}
}%
\providecommand \@ifnum [1]{%
 \ifnum #1\expandafter \@firstoftwo
 \else \expandafter \@secondoftwo
 \fi
}%
\providecommand \@ifx [1]{%
 \ifx #1\expandafter \@firstoftwo
 \else \expandafter \@secondoftwo
 \fi
}%
\providecommand \natexlab [1]{#1}%
\providecommand \enquote  [1]{``#1''}%
\providecommand \bibnamefont  [1]{#1}%
\providecommand \bibfnamefont [1]{#1}%
\providecommand \citenamefont [1]{#1}%
\providecommand \href@noop [0]{\@secondoftwo}%
\providecommand \href [0]{\begingroup \@sanitize@url \@href}%
\providecommand \@href[1]{\@@startlink{#1}\@@href}%
\providecommand \@@href[1]{\endgroup#1\@@endlink}%
\providecommand \@sanitize@url [0]{\catcode `\\12\catcode `\$12\catcode
  `\&12\catcode `\#12\catcode `\^12\catcode `\_12\catcode `\%12\relax}%
\providecommand \@@startlink[1]{}%
\providecommand \@@endlink[0]{}%
\providecommand \url  [0]{\begingroup\@sanitize@url \@url }%
\providecommand \@url [1]{\endgroup\@href {#1}{\urlprefix }}%
\providecommand \urlprefix  [0]{URL }%
\providecommand \Eprint [0]{\href }%
\providecommand \doibase [0]{http://dx.doi.org/}%
\providecommand \selectlanguage [0]{\@gobble}%
\providecommand \bibinfo  [0]{\@secondoftwo}%
\providecommand \bibfield  [0]{\@secondoftwo}%
\providecommand \translation [1]{[#1]}%
\providecommand \BibitemOpen [0]{}%
\providecommand \bibitemStop [0]{}%
\providecommand \bibitemNoStop [0]{.\EOS\space}%
\providecommand \EOS [0]{\spacefactor3000\relax}%
\providecommand \BibitemShut  [1]{\csname bibitem#1\endcsname}%
\let\auto@bib@innerbib\@empty
\bibitem [{\citenamefont {Xiao}\ \emph {et~al.}(2012)\citenamefont {Xiao},
  \citenamefont {Liu}, \citenamefont {Feng}, \citenamefont {Xu},\ and\
  \citenamefont {Yao}}]{Xiao2012}%
  \BibitemOpen
  \bibfield  {author} {\bibinfo {author} {\bibfnamefont {D.}~\bibnamefont
  {Xiao}}, \bibinfo {author} {\bibfnamefont {G.-B.}\ \bibnamefont {Liu}},
  \bibinfo {author} {\bibfnamefont {W.}~\bibnamefont {Feng}}, \bibinfo {author}
  {\bibfnamefont {X.}~\bibnamefont {Xu}}, \ and\ \bibinfo {author}
  {\bibfnamefont {W.}~\bibnamefont {Yao}},\ }\href {\doibase
  10.1103/PhysRevLett.108.196802} {\bibfield  {journal} {\bibinfo  {journal}
  {Phys. Rev. Lett.}\ }\textbf {\bibinfo {volume} {108}},\ \bibinfo {pages}
  {196802} (\bibinfo {year} {2012})}\BibitemShut {NoStop}%
\bibitem [{\citenamefont {Qiao}\ \emph {et~al.}(2012)\citenamefont {Qiao},
  \citenamefont {Jiang}, \citenamefont {Li}, \citenamefont {Yao},\ and\
  \citenamefont {Niu}}]{Qiao2012}%
  \BibitemOpen
  \bibfield  {author} {\bibinfo {author} {\bibfnamefont {Z.}~\bibnamefont
  {Qiao}}, \bibinfo {author} {\bibfnamefont {H.}~\bibnamefont {Jiang}},
  \bibinfo {author} {\bibfnamefont {X.}~\bibnamefont {Li}}, \bibinfo {author}
  {\bibfnamefont {Y.}~\bibnamefont {Yao}}, \ and\ \bibinfo {author}
  {\bibfnamefont {Q.}~\bibnamefont {Niu}},\ }\href {\doibase
  10.1103/PhysRevB.85.115439} {\bibfield  {journal} {\bibinfo  {journal} {Phys.
  Rev. B}\ }\textbf {\bibinfo {volume} {85}},\ \bibinfo {pages} {115439}
  (\bibinfo {year} {2012})}\BibitemShut {NoStop}%
\bibitem [{\citenamefont {Bergh\"auser}\ and\ \citenamefont
  {Malic}(2014)}]{Berghauser2014}%
  \BibitemOpen
  \bibfield  {author} {\bibinfo {author} {\bibfnamefont {G.}~\bibnamefont
  {Bergh\"auser}}\ and\ \bibinfo {author} {\bibfnamefont {E.}~\bibnamefont
  {Malic}},\ }\href {\doibase 10.1103/PhysRevB.89.125309} {\bibfield  {journal}
  {\bibinfo  {journal} {Phys. Rev. B}\ }\textbf {\bibinfo {volume} {89}},\
  \bibinfo {pages} {125309} (\bibinfo {year} {2014})}\BibitemShut {NoStop}%
\bibitem [{\citenamefont {Taghizadeh}\ \emph {et~al.}(2017)\citenamefont
  {Taghizadeh}, \citenamefont {Hipolito},\ and\ \citenamefont
  {Pedersen}}]{Taghizadeh2017}%
  \BibitemOpen
  \bibfield  {author} {\bibinfo {author} {\bibfnamefont {A.}~\bibnamefont
  {Taghizadeh}}, \bibinfo {author} {\bibfnamefont {F.}~\bibnamefont
  {Hipolito}}, \ and\ \bibinfo {author} {\bibfnamefont {T.~G.}\ \bibnamefont
  {Pedersen}},\ }\href {\doibase 10.1103/PhysRevB.96.195413} {\bibfield
  {journal} {\bibinfo  {journal} {Phys. Rev. B}\ }\textbf {\bibinfo {volume}
  {96}},\ \bibinfo {pages} {195413} (\bibinfo {year} {2017})}\BibitemShut
  {NoStop}%
\bibitem [{\citenamefont {Taghizadeh}\ and\ \citenamefont
  {Pedersen}(2018)}]{Taghizadeh2018}%
  \BibitemOpen
  \bibfield  {author} {\bibinfo {author} {\bibfnamefont {A.}~\bibnamefont
  {Taghizadeh}}\ and\ \bibinfo {author} {\bibfnamefont {T.~G.}\ \bibnamefont
  {Pedersen}},\ }\href {\doibase 10.1103/PhysRevB.97.205432} {\bibfield
  {journal} {\bibinfo  {journal} {Phys. Rev. B}\ }\textbf {\bibinfo {volume}
  {97}},\ \bibinfo {pages} {205432} (\bibinfo {year} {2018})}\BibitemShut
  {NoStop}%
\bibitem [{\citenamefont {Schliemann}\ and\ \citenamefont
  {Loss}(2004)}]{Schliemann2004}%
  \BibitemOpen
  \bibfield  {author} {\bibinfo {author} {\bibfnamefont {J.}~\bibnamefont
  {Schliemann}}\ and\ \bibinfo {author} {\bibfnamefont {D.}~\bibnamefont
  {Loss}},\ }\href {\doibase 10.1103/PhysRevB.69.165315} {\bibfield  {journal}
  {\bibinfo  {journal} {Phys. Rev. B}\ }\textbf {\bibinfo {volume} {69}},\
  \bibinfo {pages} {165315} (\bibinfo {year} {2004})}\BibitemShut {NoStop}%
\bibitem [{\citenamefont {Hipolito}\ \emph {et~al.}(2016)\citenamefont
  {Hipolito}, \citenamefont {Pedersen},\ and\ \citenamefont
  {Pereira}}]{Hipolito2016}%
  \BibitemOpen
  \bibfield  {author} {\bibinfo {author} {\bibfnamefont {F.}~\bibnamefont
  {Hipolito}}, \bibinfo {author} {\bibfnamefont {T.~G.}\ \bibnamefont
  {Pedersen}}, \ and\ \bibinfo {author} {\bibfnamefont {V.~M.}\ \bibnamefont
  {Pereira}},\ }\href {\doibase 10.1103/PhysRevB.94.045434} {\bibfield
  {journal} {\bibinfo  {journal} {Phys. Rev. B}\ }\textbf {\bibinfo {volume}
  {94}},\ \bibinfo {pages} {045434} (\bibinfo {year} {2016})}\BibitemShut
  {NoStop}%
\bibitem [{\citenamefont {Sipe}\ and\ \citenamefont
  {Shkrebtii}(2000)}]{Sipe2000}%
  \BibitemOpen
  \bibfield  {author} {\bibinfo {author} {\bibfnamefont {J.~E.}\ \bibnamefont
  {Sipe}}\ and\ \bibinfo {author} {\bibfnamefont {A.~I.}\ \bibnamefont
  {Shkrebtii}},\ }\href {\doibase 10.1103/PhysRevB.61.5337} {\bibfield
  {journal} {\bibinfo  {journal} {Phys. Rev. B}\ }\textbf {\bibinfo {volume}
  {61}},\ \bibinfo {pages} {5337} (\bibinfo {year} {2000})}\BibitemShut
  {NoStop}%
\bibitem [{\citenamefont {Cao}\ \emph {et~al.}(2018)\citenamefont {Cao},
  \citenamefont {Wu},\ and\ \citenamefont {Louie}}]{Cao2018}%
  \BibitemOpen
  \bibfield  {author} {\bibinfo {author} {\bibfnamefont {T.}~\bibnamefont
  {Cao}}, \bibinfo {author} {\bibfnamefont {M.}~\bibnamefont {Wu}}, \ and\
  \bibinfo {author} {\bibfnamefont {S.~G.}\ \bibnamefont {Louie}},\ }\href
  {\doibase 10.1103/PhysRevLett.120.087402} {\bibfield  {journal} {\bibinfo
  {journal} {Phys. Rev. Lett.}\ }\textbf {\bibinfo {volume} {120}},\ \bibinfo
  {pages} {087402} (\bibinfo {year} {2018})}\BibitemShut {NoStop}%
\bibitem [{\citenamefont {Cudazzo}\ \emph {et~al.}(2011)\citenamefont
  {Cudazzo}, \citenamefont {Tokatly},\ and\ \citenamefont
  {Rubio}}]{Cudazzo2011}%
  \BibitemOpen
  \bibfield  {author} {\bibinfo {author} {\bibfnamefont {P.}~\bibnamefont
  {Cudazzo}}, \bibinfo {author} {\bibfnamefont {I.~V.}\ \bibnamefont
  {Tokatly}}, \ and\ \bibinfo {author} {\bibfnamefont {A.}~\bibnamefont
  {Rubio}},\ }\href {\doibase 10.1103/PhysRevB.84.085406} {\bibfield  {journal}
  {\bibinfo  {journal} {Phys. Rev. B}\ }\textbf {\bibinfo {volume} {84}},\
  \bibinfo {pages} {085406} (\bibinfo {year} {2011})}\BibitemShut {NoStop}%
\bibitem [{\citenamefont {Wu}\ \emph {et~al.}(2015)\citenamefont {Wu},
  \citenamefont {Qu},\ and\ \citenamefont {MacDonald}}]{Wu2015}%
  \BibitemOpen
  \bibfield  {author} {\bibinfo {author} {\bibfnamefont {F.}~\bibnamefont
  {Wu}}, \bibinfo {author} {\bibfnamefont {F.}~\bibnamefont {Qu}}, \ and\
  \bibinfo {author} {\bibfnamefont {A.~H.}\ \bibnamefont {MacDonald}},\ }\href
  {\doibase 10.1103/PhysRevB.91.075310} {\bibfield  {journal} {\bibinfo
  {journal} {Phys. Rev. B}\ }\textbf {\bibinfo {volume} {91}},\ \bibinfo
  {pages} {075310} (\bibinfo {year} {2015})}\BibitemShut {NoStop}%
\bibitem [{\citenamefont {Qiu}\ \emph {et~al.}(2013)\citenamefont {Qiu},
  \citenamefont {da~Jornada},\ and\ \citenamefont {Louie}}]{Qiu2013}%
  \BibitemOpen
  \bibfield  {author} {\bibinfo {author} {\bibfnamefont {D.~Y.}\ \bibnamefont
  {Qiu}}, \bibinfo {author} {\bibfnamefont {F.~H.}\ \bibnamefont {da~Jornada}},
  \ and\ \bibinfo {author} {\bibfnamefont {S.~G.}\ \bibnamefont {Louie}},\
  }\href {\doibase 10.1103/PhysRevLett.111.216805} {\bibfield  {journal}
  {\bibinfo  {journal} {Phys. Rev. Lett.}\ }\textbf {\bibinfo {volume} {111}},\
  \bibinfo {pages} {216805} (\bibinfo {year} {2013})}\BibitemShut {NoStop}%
\bibitem [{\citenamefont {Pedersen}(2015)}]{Pedersen2015}%
  \BibitemOpen
  \bibfield  {author} {\bibinfo {author} {\bibfnamefont {T.~G.}\ \bibnamefont
  {Pedersen}},\ }\href {\doibase 10.1103/PhysRevB.92.235432} {\bibfield
  {journal} {\bibinfo  {journal} {Phys. Rev. B}\ }\textbf {\bibinfo {volume}
  {92}},\ \bibinfo {pages} {235432} (\bibinfo {year} {2015})}\BibitemShut
  {NoStop}%
\bibitem [{\citenamefont {Cadiz}\ \emph {et~al.}(2017)\citenamefont {Cadiz},
  \citenamefont {Courtade}, \citenamefont {Robert}, \citenamefont {Wang},
  \citenamefont {Shen}, \citenamefont {Cai}, \citenamefont {Taniguchi},
  \citenamefont {Watanabe}, \citenamefont {Carrere}, \citenamefont {Lagarde},
  \citenamefont {Manca}, \citenamefont {Amand}, \citenamefont {Renucci},
  \citenamefont {Tongay}, \citenamefont {Marie},\ and\ \citenamefont
  {Urbaszek}}]{Cadiz2017}%
  \BibitemOpen
  \bibfield  {author} {\bibinfo {author} {\bibfnamefont {F.}~\bibnamefont
  {Cadiz}}, \bibinfo {author} {\bibfnamefont {E.}~\bibnamefont {Courtade}},
  \bibinfo {author} {\bibfnamefont {C.}~\bibnamefont {Robert}}, \bibinfo
  {author} {\bibfnamefont {G.}~\bibnamefont {Wang}}, \bibinfo {author}
  {\bibfnamefont {Y.}~\bibnamefont {Shen}}, \bibinfo {author} {\bibfnamefont
  {H.}~\bibnamefont {Cai}}, \bibinfo {author} {\bibfnamefont {T.}~\bibnamefont
  {Taniguchi}}, \bibinfo {author} {\bibfnamefont {K.}~\bibnamefont {Watanabe}},
  \bibinfo {author} {\bibfnamefont {H.}~\bibnamefont {Carrere}}, \bibinfo
  {author} {\bibfnamefont {D.}~\bibnamefont {Lagarde}}, \bibinfo {author}
  {\bibfnamefont {M.}~\bibnamefont {Manca}}, \bibinfo {author} {\bibfnamefont
  {T.}~\bibnamefont {Amand}}, \bibinfo {author} {\bibfnamefont
  {P.}~\bibnamefont {Renucci}}, \bibinfo {author} {\bibfnamefont
  {S.}~\bibnamefont {Tongay}}, \bibinfo {author} {\bibfnamefont
  {X.}~\bibnamefont {Marie}}, \ and\ \bibinfo {author} {\bibfnamefont
  {B.}~\bibnamefont {Urbaszek}},\ }\href {\doibase 10.1103/PhysRevX.7.021026}
  {\bibfield  {journal} {\bibinfo  {journal} {Phys. Rev. X}\ }\textbf {\bibinfo
  {volume} {7}},\ \bibinfo {pages} {021026} (\bibinfo {year}
  {2017})}\BibitemShut {NoStop}%
\bibitem [{\citenamefont {Christiansen}\ \emph {et~al.}(2017)\citenamefont
  {Christiansen}, \citenamefont {Selig}, \citenamefont {Bergh\"auser},
  \citenamefont {Schmidt}, \citenamefont {Niehues}, \citenamefont {Schneider},
  \citenamefont {Arora}, \citenamefont {de~Vasconcellos}, \citenamefont
  {Bratschitsch}, \citenamefont {Malic},\ and\ \citenamefont
  {Knorr}}]{Christiansen2017}%
  \BibitemOpen
  \bibfield  {author} {\bibinfo {author} {\bibfnamefont {D.}~\bibnamefont
  {Christiansen}}, \bibinfo {author} {\bibfnamefont {M.}~\bibnamefont {Selig}},
  \bibinfo {author} {\bibfnamefont {G.}~\bibnamefont {Bergh\"auser}}, \bibinfo
  {author} {\bibfnamefont {R.}~\bibnamefont {Schmidt}}, \bibinfo {author}
  {\bibfnamefont {I.}~\bibnamefont {Niehues}}, \bibinfo {author} {\bibfnamefont
  {R.}~\bibnamefont {Schneider}}, \bibinfo {author} {\bibfnamefont
  {A.}~\bibnamefont {Arora}}, \bibinfo {author} {\bibfnamefont {S.~M.}\
  \bibnamefont {de~Vasconcellos}}, \bibinfo {author} {\bibfnamefont
  {R.}~\bibnamefont {Bratschitsch}}, \bibinfo {author} {\bibfnamefont
  {E.}~\bibnamefont {Malic}}, \ and\ \bibinfo {author} {\bibfnamefont
  {A.}~\bibnamefont {Knorr}},\ }\href {\doibase 10.1103/PhysRevLett.119.187402}
  {\bibfield  {journal} {\bibinfo  {journal} {Phys. Rev. Lett.}\ }\textbf
  {\bibinfo {volume} {119}},\ \bibinfo {pages} {187402} (\bibinfo {year}
  {2017})}\BibitemShut {NoStop}%
\bibitem [{\citenamefont {Rasmussen}\ and\ \citenamefont
  {Thygesen}(2015)}]{Rasmussen2015}%
  \BibitemOpen
  \bibfield  {author} {\bibinfo {author} {\bibfnamefont {F.~A.}\ \bibnamefont
  {Rasmussen}}\ and\ \bibinfo {author} {\bibfnamefont {K.~S.}\ \bibnamefont
  {Thygesen}},\ }\href {\doibase 10.1021/acs.jpcc.5b02950} {\bibfield
  {journal} {\bibinfo  {journal} {J. Phys. Chem. C}\ }\textbf {\bibinfo
  {volume} {119}},\ \bibinfo {pages} {13169} (\bibinfo {year}
  {2015})}\BibitemShut {NoStop}%
\bibitem [{\citenamefont {Olsen}\ \emph {et~al.}(2016)\citenamefont {Olsen},
  \citenamefont {Latini}, \citenamefont {Rasmussen},\ and\ \citenamefont
  {Thygesen}}]{Olsen2016}%
  \BibitemOpen
  \bibfield  {author} {\bibinfo {author} {\bibfnamefont {T.}~\bibnamefont
  {Olsen}}, \bibinfo {author} {\bibfnamefont {S.}~\bibnamefont {Latini}},
  \bibinfo {author} {\bibfnamefont {F.}~\bibnamefont {Rasmussen}}, \ and\
  \bibinfo {author} {\bibfnamefont {K.~S.}\ \bibnamefont {Thygesen}},\ }\href
  {\doibase 10.1103/PhysRevLett.116.056401} {\bibfield  {journal} {\bibinfo
  {journal} {Phys. Rev. Lett.}\ }\textbf {\bibinfo {volume} {116}},\ \bibinfo
  {pages} {056401} (\bibinfo {year} {2016})}\BibitemShut {NoStop}%
\bibitem [{\citenamefont {Scuri}\ \emph {et~al.}(2018)\citenamefont {Scuri},
  \citenamefont {Zhou}, \citenamefont {High}, \citenamefont {Wild},
  \citenamefont {Shu}, \citenamefont {De~Greve}, \citenamefont {Jauregui},
  \citenamefont {Taniguchi}, \citenamefont {Watanabe}, \citenamefont {Kim},
  \citenamefont {Lukin},\ and\ \citenamefont {Park}}]{Scuri2018}%
  \BibitemOpen
  \bibfield  {author} {\bibinfo {author} {\bibfnamefont {G.}~\bibnamefont
  {Scuri}}, \bibinfo {author} {\bibfnamefont {Y.}~\bibnamefont {Zhou}},
  \bibinfo {author} {\bibfnamefont {A.~A.}\ \bibnamefont {High}}, \bibinfo
  {author} {\bibfnamefont {D.~S.}\ \bibnamefont {Wild}}, \bibinfo {author}
  {\bibfnamefont {C.}~\bibnamefont {Shu}}, \bibinfo {author} {\bibfnamefont
  {K.}~\bibnamefont {De~Greve}}, \bibinfo {author} {\bibfnamefont {L.~A.}\
  \bibnamefont {Jauregui}}, \bibinfo {author} {\bibfnamefont {T.}~\bibnamefont
  {Taniguchi}}, \bibinfo {author} {\bibfnamefont {K.}~\bibnamefont {Watanabe}},
  \bibinfo {author} {\bibfnamefont {P.}~\bibnamefont {Kim}}, \bibinfo {author}
  {\bibfnamefont {M.~D.}\ \bibnamefont {Lukin}}, \ and\ \bibinfo {author}
  {\bibfnamefont {H.}~\bibnamefont {Park}},\ }\href {\doibase
  10.1103/PhysRevLett.120.037402} {\bibfield  {journal} {\bibinfo  {journal}
  {Phys. Rev. Lett.}\ }\textbf {\bibinfo {volume} {120}},\ \bibinfo {pages}
  {037402} (\bibinfo {year} {2018})}\BibitemShut {NoStop}%
\bibitem [{\citenamefont {Steinleitner}\ \emph {et~al.}(2017)\citenamefont
  {Steinleitner}, \citenamefont {Merkl}, \citenamefont {Nagler}, \citenamefont
  {Mornhinweg}, \citenamefont {Sch\"{u}ller}, \citenamefont {Korn},
  \citenamefont {Chernikov},\ and\ \citenamefont {Huber}}]{Steinleitner2017}%
  \BibitemOpen
  \bibfield  {author} {\bibinfo {author} {\bibfnamefont {P.}~\bibnamefont
  {Steinleitner}}, \bibinfo {author} {\bibfnamefont {P.}~\bibnamefont {Merkl}},
  \bibinfo {author} {\bibfnamefont {P.}~\bibnamefont {Nagler}}, \bibinfo
  {author} {\bibfnamefont {J.}~\bibnamefont {Mornhinweg}}, \bibinfo {author}
  {\bibfnamefont {C.}~\bibnamefont {Sch\"{u}ller}}, \bibinfo {author}
  {\bibfnamefont {T.}~\bibnamefont {Korn}}, \bibinfo {author} {\bibfnamefont
  {A.}~\bibnamefont {Chernikov}}, \ and\ \bibinfo {author} {\bibfnamefont
  {R.}~\bibnamefont {Huber}},\ }\href {\doibase 10.1021/acs.nanolett.6b04422}
  {\bibfield  {journal} {\bibinfo  {journal} {Nano Lett.}\ }\textbf {\bibinfo
  {volume} {17}},\ \bibinfo {pages} {1455} (\bibinfo {year}
  {2017})}\BibitemShut {NoStop}%
\end{thebibliography}%


%

\end{document}



\title{\underline{\textbf{Supplementary Material}} \\
\textbf{Nonlinear excitonic spin Hall effect in monolayer transition metal dichalcogenides}}


\author{Alireza Taghizadeh}
\email{ata@nano.aau.dk}
\affiliation{Department of Physics and Nanotechnology, Aalborg University, DK-9220 Aalborg {\O}st, Denmark}


\author{T. G. Pedersen}
\affiliation{Department of Physics and Nanotechnology, Aalborg University, DK-9220 Aalborg {\O}st, Denmark}
\affiliation{Center for Nanostructured Graphene (CNG), DK-9220 Aalborg {\O}st, Denmark}

\begin{abstract}
	This document provides supplementary information to ``Nonlinear excitonic spin Hall effect in monolayer transition metal dichalcogenides'', in which we provide the detailed derivation of various formulas as well as the numerical parameters used to compute the optical response. In addition, the OR conductivities calculated for monolayer WSe$_2$ are presented.
\end{abstract}

\maketitle

\section{A) Trigonal warping Hamiltonian}
In this section, we review the steps for deriving the TW Hamiltonian. We start by constructing a minimal tight-binding (TB) Hamiltonian for an unperturbed monolayer TMD in the basis of $|d_{z^2}\rangle$ and $(|d_{x^2-y^2}\rangle+|d_{xy}\rangle)/\sqrt{2}$ orbitals of the metal atom \cite{Xiao2012}. Including the intrinsic SOC, the Hamiltonian in $\va{k}$-space reads \cite{Qiao2012, Berghauser2014}
\begin{equation}
	\label{eq:TBHamiltonian}
	H_0(\va{k}) = {\begin{bmatrix} \dfrac{\Delta}{2}+\dfrac{\lambda}{2}g(\va{k})s & -\gamma f(\va{k}) \\ -\gamma f^*(\va{k}) & -\dfrac{\Delta}{2}-\dfrac{\lambda}{2}g(\va{k})s \end{bmatrix}} \, ,
\end{equation}
where $\Delta/2$, $\gamma$ and $\lambda$ are the on-site energy, effective hopping and SOC strength, respectively. 
The wavevector-dependent functions are given by
\begin{subequations}
	\begin{align}
	&f(\va{k}) \equiv \exp(ik_x a_0/\sqrt{3}) + 2 \exp(-ik_x a_0/2\sqrt{3}) \cos(k_ya_0/2) \, , \\
	&g(\va{k}) \equiv -4\sin(k_y a_0/2) \cos(k_x a_0/\sqrt{3})+ 2\sin(k_y a_0) \, ,
	\end{align}
\end{subequations}
with the lattice constant $a_0$.
The eigenenergies and eigenvectors of the 2-band TB Hamiltonian read
\begin{subequations}
	\begin{align}
	&\varepsilon_{s,c\va{k}} = \varepsilon \, , \quad \varepsilon_{s,v\va{k}} \equiv -\varepsilon \, , \quad \varepsilon = \sqrt{\Delta_s^2/4+\gamma^2 F^2}  \, ,  \\
	&|s,c\va{k} \rangle = [\cos(\xi/2), -\sin(\xi/2) e^{-i\phi}] \, , \quad |s,v\va{k} \rangle = [\sin(\xi/2) e^{i\phi}, \cos(\xi/2)] \, ,
	\end{align}
\end{subequations}
where $F$ and $\phi$ are defined using $f(\va{k})=F \exp(i\phi)$, and $\xi$ and $\Delta_s$ are given by $2\cos(\xi)=\Delta_s/\varepsilon$ and $\Delta_s=\Delta+s \lambda g(\va{k})$, respectively. Neglecting the intra-atomic contributions, the momentum operator in the TB method is given by the $\va{k}$-derivative of the Hamiltonian, \ie $\hbar\va{p}=m \nabla_{\va{k}} H_0(\va{k})$. 

The massive Dirac Hamiltonian can be derived from Eq.~(\ref{eq:TBHamiltonian}) by a Taylor expansion of $f(\va{k})$ and $g(\va{k})$ around the Dirac points, K: $2\pi(3^{-1/2},3^{-1})/a_0$ and K$'$: $2\pi(3^{-1/2},-3^{-1})/a_0$, to linear order in $\va{k}$.
In contrast, the TW Hamiltonian is obtained by collecting the terms up to second order in $\va{k}$ for $f(\va{k})$,
\begin{align}
	f(\va{k})|_{\textrm{K}/\textrm{K}'} \approx \dfrac{\sqrt{3}}{2} e^{i\pi/3} \big[-i(\kappa_x-i\kappa_y\tau) + \zeta(\kappa_x+i\kappa_y\tau)^2 \big] \, .
\end{align}
Here, $\kappa_\alpha\equiv a_0(k_\alpha-K_\alpha)$ is the normalized wavevector measured with respect to K or K$'$, and $\zeta=\sqrt{3}/12$ multiplies the trigonal warping term [note that $g(\va{k})|_{\textrm{K}/\textrm{K}'} \approx 3\sqrt{3}\tau$]. To calculate the optical response analytically, we expand the eigenenergies $\varepsilon_{s,v\va{k}}$ and $\varepsilon_{s,c\va{k}}$, momentum matrix elements $p_{mn}^\alpha\equiv \langle s,m\va{k}|\hat{p}_\alpha|s,n\va{k} \rangle=(p_{nm}^{\alpha})^*$ and Berry connections $\Omega_{nm}^\alpha\equiv\langle s,n\va{k}|i\partial_{k_\alpha}| s,m\va{k} \rangle$ to first order in $\zeta$. For instance, the expression for the eigenenergies reads
\begin{align}
	&\varepsilon_{s,c\va{k}} = -\varepsilon_{s,v\va{k}} \approx \varepsilon_0 \big[1- \zeta \tau \kappa (1-\delta_{s\tau}^2) \sin(3\theta) \big] \, , 
\end{align}
where $2\varepsilon_0\equiv\sqrt{\Delta_{s\tau}^2+3\gamma^2\kappa^2}$ and $2\delta_{s\tau}\equiv\Delta_{s\tau}/\varepsilon_0$ with $\Delta_{s\tau}\equiv\Delta+3\sqrt{3}\lambda s\tau$. $\kappa$ and $\theta$ are the magnitude and phase of the normalized wavevector $\boldsymbol{\kappa}$. The expressions for other required parameters can be determined similarly.

\section{B) Second-order Optical conductivity}
In this section, we explain the details of determining the second-order charge and spin conductivities.
The light-matter interaction can be studied by solving the master equation for the density matrix, $i\hbar\partial_t \hat{\rho}=[\hat{H},\hat{\rho}]+i\hat{\mathcal{L}}(\hat{\rho})$, where $\hat{H}$ and $\hat{\mathcal{L}}(\hat{\rho})$ are the total Hamiltonian and Lindblad superoperator, respectively. The total Hamiltonian consists of an unperturbed part (free electron plus the electron-hole interaction) and a light-matter interaction part. In the IPA limit, the electron-hole part is ignored, whereas it is treated in the mean-field approximation when excitonic effects are included. 
Moreover, the interaction Hamiltonian in the dipole approximation (long-wavelength regime) reads $\hat{H}_\textrm{int}(t)=e\hat{\va{r}}\vdot\field{E}(t)$, where $\hat{\va{r}}$ and $\field{E}(t)$ are the position operator and time-dependent electric field, respectively.
Finally, the Lindbald superoperator is evaluated in the context of the relaxation-time approximation using phenomenological broadening parameters, rather than the exact calculation. We use two different relaxation rates: $\Gamma_e$ for the coherences ($\rho_{cv\va{k}}$, $\rho_{vc\va{k}}$) and $\Gamma_i$ for the band populations ($\rho_{cc\va{k}}$, $\rho_{vv\va{k}}$).

The master equation is solved perturbatively up to any required order in the external field, and the $n$th-order density matrix elements $\rho_{mn\va{k}}^{(n)}$ are determined. Thereafter, the $n$th-order charge and spin current densities are evaluated by $\va{J}^{(nC)}=\textrm{Tr}[\hat{\rho}^{(n)}\hat{\va{j}}_c]$ and $\va{J}^{(nS)}=\textrm{Tr}[\hat{\rho}^{(n)}\hat{\va{j}}_s]$, respectively. The charge current density operator reads $\hat{\va{j}}_c=-e\hat{\va{v}}/A$, in which $\hat{\va{v}}$ and $A$ are the velocity operator and crystal area, respectively \cite{Taghizadeh2017, Taghizadeh2018}. The spin current density operator for spin moment polarized along the $z$-direction (perpendicular to the TMD plane) is given by $\hat{\va{j}}_s=(\hat{s}_z\hat{\va{j}}_c+\hat{\va{j}}_c\hat{s}_z)/2$ \cite{Schliemann2004}. Here, $\hat{s}_z$ is the $z$ component of the spin operator with the eigenvalues $s=\pm1$. With this definition, the dimension of the spin current density is the same as the charge current density. 

If the incident electric field is decomposed into its harmonic components, \ie $\field{E}(t)=\sum_{\omega_1} \field{E}_{\omega_1} \exp(-i\omega_1t)$, the second-order charge/spin current density reads
\begin{align}
	&J_{\eta}^{(2X)}(t)=\sum_{\omega_1,\omega_2} \sum_{\alpha,\beta} \sigma_{\eta\alpha\beta}^{(2X)}(\omega_1+\omega_2) \mathcal{E}_{\omega_1}^\alpha \mathcal{E}_{\omega_2}^\beta e^{-i(\omega_1+\omega_2)t} \, , 
\end{align}
where $\sigma_{\eta\alpha\beta}^{(2X)}(\omega_1+\omega_2)$ is the second-order charge/spin conductivity tensor with $X=C$/$S$. Note that the summations over $\omega_1$ and $\omega_2$ include both positive and negative frequencies. Due to the point-group symmetry of the honeycomb lattice, there can be only two independent tensor components for the quadratic conductivities, \ie $-\sigma_{\alpha\alpha\alpha}^{(2X)}=\sigma_{\alpha\beta\beta}^{(2X)}=\sigma_{\beta\alpha\beta}^{(2X)}=\sigma_{\beta\beta\alpha}^{(2X)}$, where $(\alpha,\beta)$ are $(x,y)$ or $(y,x)$ \cite{Hipolito2016}. Assuming a monochromatic field with an arbitrary polarization, \ie $\field{E}(t)=(\mathcal{E}_x\va{e}_x+\mathcal{E}_y\va{e}_y)\exp(-i\omega t)+ \mathrm{c.c.}$, the total quadratic current density reads $\va{J}^{(2X)}(t)=\va{J}_0^{(2X)}+\va{J}_{2\omega}^{(2X)} \exp(-2i\omega t)  + \mathrm{c.c.}$. Here, $\va{J}_0^{(2X)}$ and $\va{J}_{2\omega}^{(2X)}$ are the induced OR and SHG current densities, respectively, given by
\begin{subequations}
\label{eq:Circular}
\begin{align}
\label{eq:CircularA}
	&\begin{bmatrix} J_{0,x}^{(2X)} \\ J_{0,y}^{(2X)} \end{bmatrix} = 
	\begin{bmatrix}
	\sigma_{xxx}^{(2X)}(\omega-\omega) && -\sigma_{yyy}^{(2X)}(\omega-\omega) \\
	-\sigma_{yyy}^{(2X)}(\omega-\omega) && -\sigma_{xxx}^{(2X)}(\omega-\omega)
	\end{bmatrix} 
	\begin{bmatrix}
	|\mathcal{E}_x|^2-|\mathcal{E}_y|^2 \\ 2\Re{\mathcal{E}_x\mathcal{E}_y^*}
	\end{bmatrix} \, , \\
	\label{eq:CircularB}
	&\begin{bmatrix} J_{2\omega,x}^{(2X)} \\ J_{2\omega,y}^{(2X)} \end{bmatrix} = 
	\begin{bmatrix}
	\sigma_{xxx}^{(2X)}(\omega+\omega) && -\sigma_{yyy}^{(2X)}(\omega+\omega) \\
	-\sigma_{yyy}^{(2X)}(\omega+\omega) && -\sigma_{xxx}^{(2X)}(\omega+\omega)
	\end{bmatrix} 
	\begin{bmatrix} \mathcal{E}_x^2-\mathcal{E}_y^2 \\ 2\mathcal{E}_x\mathcal{E}_y \end{bmatrix} \, .
\end{align}
\end{subequations}
In the absence of external magnetic field, $\sigma_{yyy}^{(2C)}$ and $\sigma_{xxx}^{(2S)}$ vanish due to TRS as already discussed in the main text.
Hence, using Eq.~(\ref{eq:CircularA}), it can be shown that the induced OR current density vanishes for a circularly-polarized excitation, \ie $\mathcal{E}_x=\pm i\mathcal{E}_y$. Furthermore, focusing on the left circularly-polarized light, we can define the valley-dependent conductivities as 
\begin{subequations}
	\label{eq:ValleyConductivity}
	\begin{align}
	\sigma_{\mathrm{K}}^{(2C)} = \sigma_{x}^{\mathrm{A}}+\sigma_{x}^{\mathrm{B}}-i\sigma_{y}^{\mathrm{A}}-i\sigma_{y}^{\mathrm{B}} \, , \quad \sigma_{\mathrm{K}'}^{(2C)} = \sigma_{x}^{\mathrm{A}}+\sigma_{x}^{\mathrm{B}}+i\sigma_{y}^{\mathrm{A}}+i\sigma_{y}^{\mathrm{B}} \, , \\
	\sigma_{\mathrm{K}}^{(2S)} = \sigma_{x}^{\mathrm{A}}-\sigma_{x}^{\mathrm{B}}-i\sigma_{y}^{\mathrm{A}}+i\sigma_{y}^{\mathrm{B}} \, , \quad \sigma_{\mathrm{K}'}^{(2S)} = \sigma_{x}^{\mathrm{A}}-\sigma_{x}^{\mathrm{B}}+i\sigma_{y}^{\mathrm{A}}-i\sigma_{y}^{\mathrm{B}} \, ,
	\end{align}
\end{subequations}
where $\sigma_{x}^{\mathrm{A}} \equiv \sigma_{xxx}^{(2,\uparrow \mathrm{K})}=\sigma_{xxx}^{(2,\downarrow \mathrm{K}')}$, $\sigma_{x}^{\mathrm{B}} \equiv \sigma_{xxx}^{(2,\downarrow \mathrm{K})}=\sigma_{xxx}^{(2,\uparrow \mathrm{K}')}$, $\sigma_{y}^{\mathrm{A}} \equiv \sigma_{yyy}^{(2,\uparrow \mathrm{K})}=-\sigma_{yyy}^{(2,\downarrow \mathrm{K}')}$, and $\sigma_{y}^{\mathrm{B}} \equiv \sigma_{yyy}^{(2,\downarrow \mathrm{K})}=-\sigma_{yyy}^{(2,\uparrow \mathrm{K}')}$. Similarly, for a right circularly-polarized beam, the conductivity contributions of the two valleys are distinguished, as seen by exchanging K and K$'$ in Eqs.~(\ref{eq:ValleyConductivity}).

In the dipole approximation, the calculation of optical conductivity in periodic systems involves handling the ill-defined position operator. Despite the problems associated with the position operator, the optical response can be computed by formally separating the position operator into its interband and intraband parts, \ie $\hat{\va{r}}=\hat{\va{r}}^{(e)}+\hat{\va{r}}^{(i)}$ with \cite{Hipolito2016, Taghizadeh2017}
\begin{subequations}
	\label{eq:PositionOperator}
	\begin{align}
	&\va{r}_{nm}^{(e)} \equiv \mel{s,n\va{k}}{\hat{\va{r}}^{(e)}}{s,m\va{k}'} = (1-\delta_{nm}) \delta_{\va{k} \va{k}'} \va{\Omega}_{nm} \, , \\
	&\va{r}_{nm}^{(i)} \equiv \mel{s,n\va{k}}{\hat{\va{r}}^{(i)}}{s,m\va{k}'} = \delta_{nm} \big( \va{\Omega}_{nn} + i\gradient_{\va{k}} \big) \delta_{\va{k}\va{k}'} \, .
	\end{align}
\end{subequations}
The intraband part of the position operator leads to the appearance of the so-called generalized derivative as discussed below.
For the quadratic optical response, four different combinations of interband ($e$) and intraband ($i$) terms, denoted by $ee$, $ie$, $ei$ and $ii$, are obtained \cite{Hipolito2016, Taghizadeh2017}. At zero temperature, when the Fermi level resides in the middle of bandgap, the $ei$ and $ii$ terms vanish \cite{Taghizadeh2017}. Therefore, the expressions for charge and spin conductivities read $\sigma_{\eta\alpha\beta}^{(2C)}(\omega_1+\omega_2) \equiv \sum_s \big[\sigma_{s,\eta\alpha\beta}^{(2,ee)}(\omega_1+\omega_2) + \sigma_{s,\eta\alpha\beta}^{(2,ie)}(\omega_1+\omega_2)\big]$, and $\sigma_{\eta\alpha\beta}^{(2C)}(\omega_1+\omega_2) \equiv \sum_s s \big[\sigma_{s,\eta\alpha\beta}^{(2,ee)}(\omega_1+\omega_2) + \sigma_{s,\eta\alpha\beta}^{(2,ie)}(\omega_1+\omega_2)\big]$, respectively. Here, $\sigma_{s,\eta\alpha\beta}^{(2,ee)}$ and $\sigma_{s,\eta\alpha\beta}^{(2,ie)}$ are the purely interband and mixed intraband-interband part of the conductivities for each spin, respectively, which are simply referred to as interband and intraband contributions. For a two-band system, the interband part, $\sigma_{s,\eta\alpha\beta}^{(2,ee)}$, originates from the band populations, $\rho_{cc\va{k}}^{(2)}$ and $\rho_{vv\va{k}}^{(2)}$, whereas the intraband term, $\sigma_{s,\eta\alpha\beta}^{(2,ie)}$, emerges from the coherences, $\rho_{cv\va{k}}^{(2)}$ and $\rho_{vc\va{k}}^{(2)}$.

\subsection{Independent particle approximation}
If the electron-hole interaction is ignored, the expressions for inter- and intraband conductivities of each spin, assuming a single valence and conduction band per spin, are given by \cite{Taghizadeh2017}
\begin{subequations}
	\label{eq:IPAConductivity}
	\begin{align}
	\sigma_{s,\eta\alpha\beta}^{(2,ee)} \equiv &+C_{0} \sum_{\va{k}} \dfrac{\dd \varepsilon_{cv}/\dd k_\eta}{\varepsilon_{cv}^2(\hbar\omega_1+\hbar\omega_2+i\Gamma_i)} \Big[ \dfrac{p_{vc}^\alpha p_{cv}^\beta}{\hbar\omega_1+i\Gamma_e+\varepsilon_{cv}}-\dfrac{p_{cv}^\alpha p_{vc}^\beta}{\hbar\omega_1+i\Gamma_e-\varepsilon_{cv}}\Big] + (\omega_1 \rightleftarrows \omega_2) \, , \\
	\sigma_{s,\eta\alpha\beta}^{(2,ie)} \equiv &- C_{0} \sum_{\va{k}} \dfrac{p_{vc}^\eta}{\hbar\omega_1+\hbar\omega_2+i\Gamma_e-\varepsilon_{cv}} \bigg[ \dfrac{p_{cv}^{\alpha}}{\varepsilon_{cv}(\hbar\omega_1+i\Gamma_e-\varepsilon_{cv})} \bigg]_{;k_\beta} \nonumber \\ 
	&-C_{0} \sum_{\va{k}} \dfrac{p_{cv}^\eta}{\hbar\omega_1+\hbar\omega_2+i\Gamma_e+\varepsilon_{cv}} \bigg[ \dfrac{p_{vc}^{\alpha}}{\varepsilon_{cv}(\hbar\omega_1+i\Gamma_e+\varepsilon_{cv})} \bigg]_{;k_\beta} + (\omega_1 \rightleftarrows \omega_2) \, . 
	\end{align}
\end{subequations}
Here, $C_{0} \equiv \dfrac{e^3 \hbar}{2m^2 A}$, $\varepsilon_{cv} \equiv \varepsilon_{s,c\va{k}}-\varepsilon_{s,v\va{k}}$ and $p_{cv}^\alpha$ are the transition energy and interband momentum matrix element for each spin, respectively. In addition, $[o_{nm}]_{;k_\alpha}\equiv \dd o_{nm}/\dd k_\alpha-i(\Omega_{nn}^\alpha-\Omega_{mm}^\alpha)o_{nm}$ denotes the generalized derivative \cite{Taghizadeh2017}, including the Berry connection. Note that the summation over $\va{k}$ implies an integral over the  Brillouin zone (BZ), \ie $(2\pi)^2 \sum_\va{k} \rightarrow A \int_{\mathrm{BZ}} \dd[2]{\va{k}}$. Furthermore, the conductivity expressions are symmetrized with respect to the frequencies, and are valid for any values of $\omega_1$ and $\omega_2$. Since the generalized derivative obeys the ordinary derivative chain rule, the intraband term can be rewritten by using 
\begin{align}
	\label{eq:ChainRule}
	&\bigg[ \dfrac{p_{cv}^{\alpha}}{\varepsilon_{cv}(\hbar\omega_1+i\Gamma_e-\varepsilon_{cv})} \bigg]_{;k_\beta} = \dfrac{\big[p_{cv}^{\alpha}\big]_{;k_\beta}}{\varepsilon_{cv}(\hbar\omega_1+i\Gamma_e-\varepsilon_{cv})} -\dfrac{p_{cv}^{\alpha}(\hbar\omega_1+i\Gamma_e-2\varepsilon_{cv})}{\varepsilon_{cv}^2(\hbar\omega_1+i\Gamma_e-\varepsilon_{cv})^2} \dv{\varepsilon_{cv}}{k_\beta} \, ,  
\end{align}
where we have used the fact that $[\varepsilon_{cv}]_{;k_\beta}=\dd{\varepsilon_{cv}}/\dd{k_\beta}$. For the quadratic charge conductivity, the contribution of the last term in Eq.~(\ref{eq:ChainRule}) vanishes due to TRS. Analogously, the $ee$ term does not contribute to the quadratic charge conductivity. In contrast, for the second-order spin conductivities, both terms involving the energy derivatives survive and contribute to the total optical response.

For the quadratic charge conductivity, the so-called shift current emerges from the first term on the right-hand side of Eq.~(\ref{eq:ChainRule}), which includes the generalized derivative. To illustrate this fact, we consider the case of $\alpha=\eta$, and write the contribution due to this term as
\begin{align}
	\label{eq:ShiftVector}
	&\sum_{\va{k}} \dfrac{p_{vc}^\alpha \big[p_{cv}^{\alpha}\big]_{;k_\beta}}{d(\va{k})} = \sum_{\va{k}} \Big[ \dfrac{\abs{p_{cv}^\alpha}}{d(\va{k})} \pdv{\abs{p_{cv}^\alpha}}{k_\beta} + i\dfrac{\abs{p_{cv}^\alpha}^2 R_{cv}^{\alpha,\beta}}{d(\va{k})} \Big] \quad \xRightarrow{\quad\mathrm{TRS}\quad} \quad \sum_{\va{k}} i\dfrac{\abs{p_{cv}^\alpha}^2 R_{cv}^{\alpha,\beta}}{d(\va{k})} \, . 
\end{align}
Here, $d(\va{k}) \equiv \varepsilon_{cv}(\hbar\omega_1+\hbar\omega_2+i\Gamma_e-\varepsilon_{cv})(\hbar\omega_1+i\Gamma_e-\varepsilon_{cv})$ is the integrand denominator (an even function of $\va{k}$). In addition, the shift vector is defined as $R_{cv}^{\alpha,\beta} \equiv \partial \phi_{cv}^\alpha/\partial k_\beta-(\Omega_{cc}^\alpha-\Omega_{vv}^\alpha)\phi_{cv}^\alpha$, in which $\phi_{cv}^\alpha$ is the phase of $p_{cv}^\alpha$, \ie $p_{cv}^\alpha=\abs{p_{cv}^\alpha}\exp(i\phi_{cv}^\alpha)$. Note that the contribution due to the derivative of the momentum matrix magnitude vanishes due to TRS, and only the shift vector term survives. Finally, taking the limit of vanishing scattering ($\Gamma_e\rightarrow 0$) and letting $\omega_1=-\omega_2=\omega$, the well-known expression of the shift current \cite{Sipe2000} is obtained.

\begin{figure}[t]
	\includegraphics[width=0.8\textwidth]{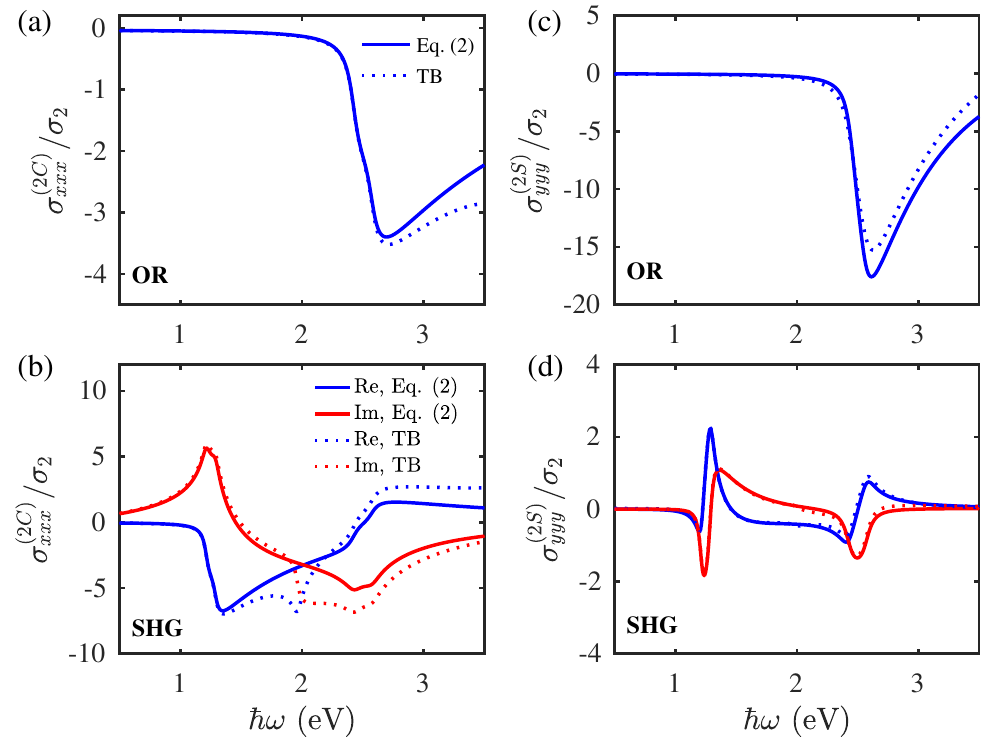}
	\caption[TB Comparison]{Quadratic charge (a,b) and spin (c,d) optical conductivities calculated for monolayer MoS$_2$ at $T=300$ K using Eq.~(2) of the main text (solid) or full TB (dotted). Two important quadratic responses are shown: the OR (a,c) and SHG (b,d). The conductivities are normalized to $\sigma_2=\SI{1e-15}{SmV^{-1}}$. The deviations at large frequencies are mainly due to the van Hove singularity transitions. }
	\label{fig:TBSpectrum}
\end{figure}


Using the TW Hamiltonian in the IPA, we can derive analytic expressions for $\sigma_{s,\eta\alpha\beta}^{(2,ee)}$ and $\sigma_{s,\eta\alpha\beta}^{(2,ie)}$, and, hence, determine $\sigma_{\eta\alpha\beta}^{(2C)}$ and $\sigma_{\eta\alpha\beta}^{(2S)}$ analytically. The expressions are provided in the main text [Eqs.~(2a) and (2b)], and the functions $\mathcal{F}_e(a,b)$ and $\mathcal{F}_i(a,b)$ are defined as
\begin{subequations}
	\begin{align}
	&\mathcal{F}_{e}(a, b) \equiv \dfrac{3\Delta_{s\tau}^3 \left(a^3+2 a b^2+b^3\right)+\Delta_{s\tau} a^2 (a+b)^3}{2a^3b(a+b)^3} -\dfrac{3\Delta_{s\tau}^4 (b-a)+2 \Delta_{s\tau}^2 a^3+a^4(a+b)}{2a^4 b^2} \nonumber \\
	&\times \tanh ^{-1}\left(\frac{a}{\Delta_{s\tau} }\right) -\frac{3 \Delta_{s\tau}^4 (a+3b)-2 \Delta_{s\tau}^2 (a+2b)(a+b)^2-(a+b)^5}{2b^2 (a+b)^4} \tanh ^{-1}\left(\frac{a+b}{\Delta_{s\tau} }\right) \, , \\
	&\mathcal{F}_{i}(a, b) \equiv -\dfrac{3\Delta_{s\tau}^3-\Delta_{s\tau} a^2}{2a^3(a+b)}+\dfrac{3\Delta_{s\tau}^4-2 \Delta_{s\tau}^2 a^2-a^4}{2a^4(a+b)} \tanh^{-1}\left(\frac{a}{\Delta_{s\tau}}\right) \, .
	\end{align}
\end{subequations}
Figures~\ref{fig:TBSpectrum}(a) and \ref{fig:TBSpectrum}(b) compare the charge and spin OR conductivities calculated for monolayer MoS$_2$ at $T=300$ K by employing Eq.~(2) with the one computed using the full TB Hamiltonian. Similarly, we show the charge and spin SHG conductivities in Figs.~\ref{fig:TBSpectrum}(c) and \ref{fig:TBSpectrum}(d). In all cases, the results obtained by employing the analytic expressions agree quite well with the full numerical solutions using the TB Hamiltonian in a wide range of frequencies. For finite value of $T$, the $ei$ and $ii$ terms are non-zero and contribute to the optical response. However, we numerically confirm that their contribution is fairly small in monolayer TMDs, since the bandgap is much larger than $k_BT$. Finally, the analytical results deviate from the TB results for frequencies close to the transitions at van Hove singularity, since the TW Hamiltonian does not capture the band structure properly in the vicinity of the M point.

\subsection{Excitonic effects}
Including the excitonic effects, the expressions for the optical conductivities read \cite{Taghizadeh2018}
\begin{subequations}
	\label{eq:ExcitonConductivity}
	\begin{align}
	\sigma_{s,\eta\alpha\beta}^{(2,ie)} &= 
	-C_{0} \sum_{n,m} \bigg[ \dfrac{\Pi_{0n}^\eta Q_{nm}^\alpha X_{m0}^\beta}{(\hbar\omega_1+\hbar\omega_2+i\Gamma_e-E_n)(\hbar\omega_1+i\Gamma_e-E_m)} + \nonumber \\
	&\qquad \qquad \qquad \dfrac{\Pi_{n0}^\eta Q_{mn}^\alpha X_{0m}^\beta}{(\hbar\omega_1+\hbar\omega_2+i\Gamma_e+E_n)(\hbar\omega_1+i\Gamma_e+E_m)} + (\omega_1 \rightleftarrows \omega_2) \bigg]  \, , \\
	\sigma_{s,\eta\alpha\beta}^{(2,ee)} &= C_{0} \dfrac{\hbar}{m} \dfrac{\hbar\omega_1+\hbar\omega_2+2i\Gamma_e}{\hbar\omega_1+\hbar\omega_2+i\Gamma_i} \sum_{n,m} \bigg[ \dfrac{X_{0n}^\eta \Pi_{nm}^\alpha X_{m0}^\beta}{(\hbar\omega_1+i\Gamma_e+E_n)(\hbar\omega_2+i\Gamma_e-E_m)} + (\omega_1 \rightleftarrows \omega_2) \bigg] \, ,
	\end{align}
\end{subequations}
where the exciton matrix elements are defined as
\begin{subequations}
	\label{eq:ExcitonMatrix}
	\begin{align}
	&X_{0n}^\alpha = (X_{n0}^\alpha)^* \equiv \dfrac{\hbar}{m} \langle 0|r_\alpha^{(e)}|\psi^{(n)} \rangle = \sum_{\va{k}} \psi_{\va{k}}^{(n)} \dfrac{p_{vc}^\alpha}{i\varepsilon_{cv}} \, , \qquad 
	\Pi_{0n}^\alpha \equiv -iE_n X_{0n}^\alpha \, , \\
	&Q_{nm}^\alpha \equiv \langle \psi^{(n)}|r_\alpha^{(i)}|\psi^{(m)} \rangle = i \sum_{\va{k}} \psi_{\va{k}}^{(n)*} \big[\psi_{\va{k}}^{(m)} \big]_{;k_\alpha} \, , \qquad \Pi_{nm}^\alpha \equiv \dfrac{m}{\hbar}i(E_n-E_m)Q_{nm}^\alpha \, . 
	\end{align} 
\end{subequations}
Here, $E_n$ and $\psi_{\va{k}}^{(n)}$ are the energy and $\va{k}$-space wavefunction of the $n$th-exciton for each spin, respectively \cite{Taghizadeh2018}. In Eq.~(\ref{eq:ExcitonMatrix}), $O_{0n}$ is the matrix element for a transition between the ground state and an exciton state, whereas $O_{nm}$ corresponds to transitions between exciton states. These expressions are modified slightly compared to the ones reported in our previous work, Ref.~\cite{Taghizadeh2018}, by including the relaxation terms. The most prominent modification is the additional frequency-dependent factor appearing in the interband term. For the SHG, this extra factor has negligible effects on the spectra at large frequencies, because $(2\omega+2i\Gamma_e)/(2\omega+i\Gamma_i) \approx 1$. In contrast, it becomes 2$\Gamma_e/\Gamma_i$ for the OR response, which may vary considerably depending on the values of the scattering rates. The temperature-dependent behavior of the excitonic OR spectra, shown in Fig.~2(a) of the main text, mainly emerges from the different variation of $\Gamma_e$ and $\Gamma_i$ with respect to temperature.

\begin{figure}[t]
	\includegraphics[width=0.8\textwidth]{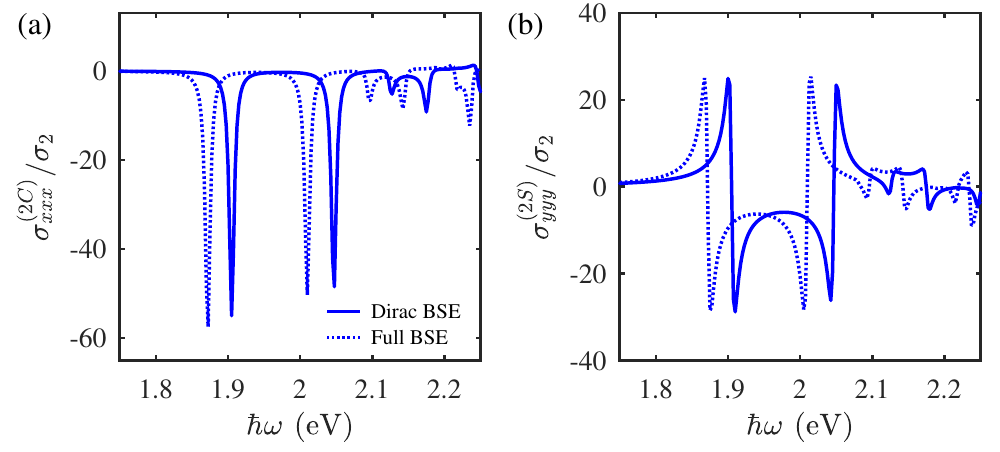}
	\caption[BSE Spectrum]{Excitonic (a) charge and (b) spin OR conductivities calculated for monolayer MoS$_2$ at $T=4$ K using the Dirac BSE (solid), see the text, and full BSE (dashed).}
	\label{fig:BSEORSpectrum}
\end{figure}
The excitonic wavefunctions and energies in Eqs.~(\ref{eq:ExcitonConductivity}) and (\ref{eq:ExcitonMatrix}) are obtained by solving the Bethe-Salpeter equation (BSE).
Ignoring the exciton center-of-mass motion due to the negligible photon momentum, the exciton state can be expressed as $|\psi^{(n)}\rangle=\frac{1}{A}\sum_\va{k}\psi_\va{k}^{(n)} |s,cv\va{k}\rangle =\frac{1}{A}\sum_\va{k}\psi_\va{k}^{(n)} \hat{c}_{c\va{k}}^\dagger \hat{c}_{v\va{k}}|0\rangle$, where $\hat{c}^\dagger$ ($\hat{c}$) is the creation (annihilation) operator, and $|0\rangle$ is the ground state (Fermi sea of electrons). The exciton wavefunction amplitude $\psi_\va{k}^{(n)}$ satisfies the BSE, which is written as \cite{Cao2018} 
\begin{align}
	(\varepsilon_{s,c\va{k}}-\varepsilon_{s,v\va{k}})\psi_\va{k}^{(n)}-\frac{1}{A}\sum_{\va{k}'} \langle s,cv\va{k}|K_{eh}|s,cv\va{k}'\rangle \psi_\va{k'}^{(n)}=E_n \psi_\va{k}^{(n)} \, .
\end{align}
Here, $K_{eh}$ is the electron-hole kernel including the direct and exchange contributions. 
We neglect the exchange part due to its minor influence \cite{Taghizadeh2018}, and use the Keldysh potential for the direct Coulomb interaction. The Keldysh potential in reals space is given by \cite{Cudazzo2011}
\begin{align}
	\mathcal{V}(\va{r}) &= \dfrac{e^2}{8\epsilon_0r_0} \bigg[ \mathbb{H}_0\Big(\frac{\epsilon_s r}{r_0}\Big) - \mathbb{Y}_0\Big(\frac{\epsilon_s r}{r_0}\Big)\bigg] \, ,
\end{align}
where $r=|\va{r}|$, and $\mathbb{H}_0$ and $\mathbb{Y}_0$ are the Struve function and Bessel function of second type, respectively. The two parameters $\epsilon_s$ and $r_0$ are the substrate screening and screening length, respectively. Using the Keldysh potential, the Coulomb matrix elements in the BSE are given by \cite{Wu2015}
\begin{align}
	\langle s,cv\va{k}|K_{eh}|s,cv\va{k}'\rangle = \dfrac{e^2}{2\epsilon_0} \dfrac{\langle s,c\va{k}|s,c\va{k}'\rangle \langle s,v\va{k}'|s,v\va{k}\rangle }{\abs{\va{k}-\va{k}'}(\epsilon_s+r_0\abs{\va{k}-\va{k}'})} \, ,
\end{align}
where $\langle s,c\va{k}|s,c\va{k}'\rangle$ and $\langle s,v\va{k}|s,v\va{k}'\rangle$ are the Bloch overlaps of conduction and valence states, respectively.

\begin{figure}[t]
	\includegraphics[width=0.8\textwidth]{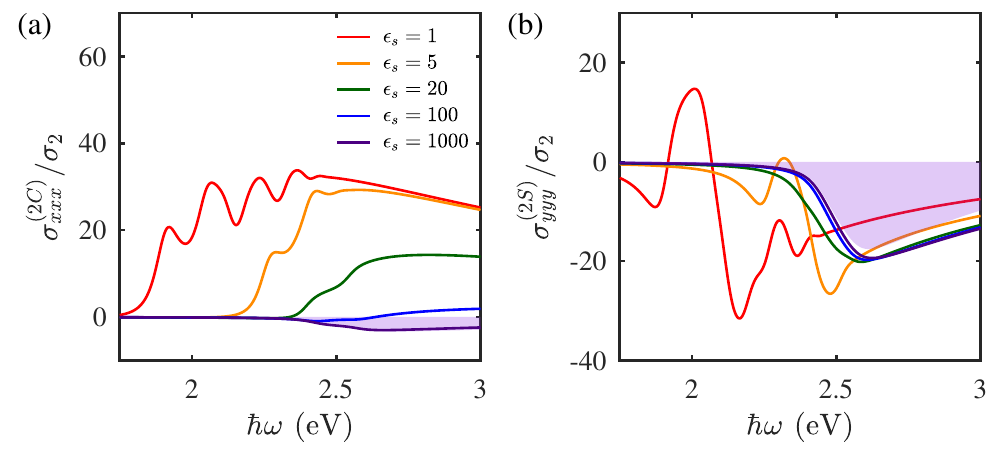}
	\caption[Screening]{Excitonic (a) charge and (b) spin OR conductivities calculated for monolayer MoS$_2$ at $T=300$ K for five different values of substrate screening, $\epsilon_s$. For comparison purposes, the IPA results are also plotted (filled purple area). }
	\label{fig:ORSpectrumScreaning}
\end{figure}

The BSE can be solved rigorously for a given quasi-particle band structure \cite{Qiu2013, Wu2015, Pedersen2015, Taghizadeh2018}. However, its approximate solution employing the massive Dirac approximation provides not only an accurate spectrum but also valuable physical information \cite{Wu2015}. For the calculation in the present work, the excitonic wavefunctions and energies, $\psi_{\va{k}}^{(n)}$ and $E_n$, are determined using the Dirac Hamiltonian, whereas the single-particles energies and momenta in Eqs.~(\ref{eq:ExcitonMatrix}) are evaluated using the TW Hamiltonian, in order to obtain a non-zero quadratic optical response. 
In Fig.~\ref{fig:BSEORSpectrum}, we compare the charge and spin OR spectra obtained using this approach with the full BSE calculations. Besides a minor energy shift of approximately 30 meV, the spectra are in very good agreement with each other. Note that the energy shift is due to the underestimation of the Coulomb potential and can be compensated merely by a slight reduction of the screening parameters. In order to validate our excitonic model, we show the charge and spin OR spectra obtained for monolayer MoS$_2$ at room temperature for five different values of substrate screening in Figs.~\ref{fig:ORSpectrumScreaning}(a) and \ref{fig:ORSpectrumScreaning}(b), respectively. For comparison purposes, we also plot the IPA results computed using the analytical expression, Eq.~(2). For both charge and spin conductivities, the excitonic spectra converge to the IPA results when the value of substrate screening is increased. 


\begin{figure}[t]
	\includegraphics[width=0.8\textwidth]{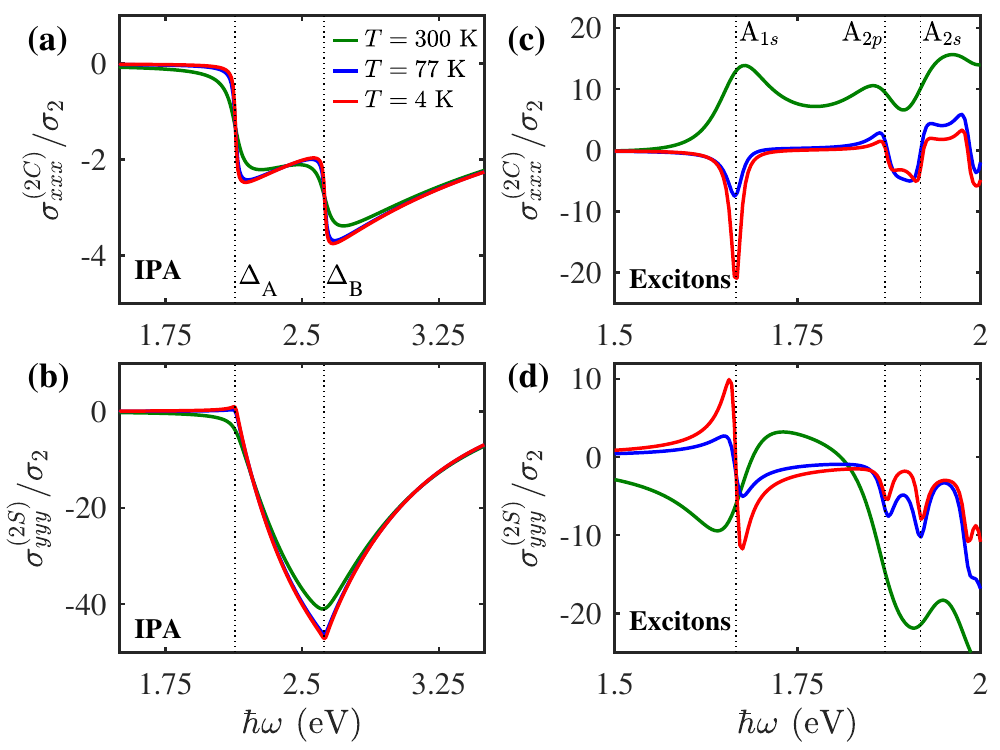}
	\caption[WSe2 Spectrum]{Charge (a,c) and spin (b,d) OR spectra calculated for monolayer WSe$_2$ at $T=300$ K (green), 77 K (blue) and 4 K (red) with the IPA model (a,b) or with excitons (c,d).  }
	\label{fig:ORSpectrumWSe2}
\end{figure}

\section{C) Numerical results and parameters}
So far, all numerical results are provided for suspended ($\epsilon_s=1$) monolayer MoS$_2$. However, other members of the TMD family are expected to behave similarly. In Fig.~\ref{fig:ORSpectrumWSe2}, we illustrate the charge and spin OR conductivities computed for monolayer WSe$_2$ at $T=4$, 77 and 300 K. The results are similar to the ones obtained for monolayer MoS$_2$, presented in Fig.~2 of the main text. Nonetheless, due to the large SOC strength in WSe$_2$, the resonances of B excitons are spectrally separated from the A resonances, \ie all features shown in Figs.~\ref{fig:ORSpectrumWSe2}(c) and \ref{fig:ORSpectrumWSe2}(d) correspond to A excitons.

Table~\ref{tab:Simulation} lists the value of all required model parameters for monolayer MoS$_2$ and WSe$_2$. The parameters $\Delta$, $\gamma$ and $\lambda$ can be determined by fitting to the experimental data or calculated quasi-particle band structures. The hopping parameter, $\gamma$, can be related to the effective mass at the Dirac points using $3\gamma^2=2\hbar^2\Delta/m_\mathrm{eff}$, where we take the average of conduction and valence band effective masses for $m_\mathrm{eff}$. For monolayer MoS$_2$, the evolution of $\Gamma_e$ with temperature is modeled using the phenomenological equation \cite{Cadiz2017}: 
\begin{align}
	\Gamma_e=\Gamma_0+c_1T+\dfrac{c_2}{\exp(\Omega/k_BT)-1} \, , 
\end{align}
where $\Gamma_0=4$ meV, $c_1=70$ $\mu$eV/K, $c_2=42.6$ meV, and $\Omega=24.2$ meV. For monolayer WSe$_2$, we use $\Gamma_e$ values at different temperatures reported in Ref.~\onlinecite{Christiansen2017}.

\begin{table}[t]
	\caption{The parameters used for generating the numerical results in the present work. The set of three values for $\Gamma_e$ are used at temperature, $T=4$/77/300 K. }
	\label{tab:Simulation}
	\centering
	\begin{tabular}{p{2cm}| P{1.7cm} P{1.7cm} P{1.7cm} P{1.7cm} P{1.7cm} P{2cm} P{2.5cm}} 
		\hline
		Material & $a_0$ (\AA) & $r_0$ (\AA) & $\Delta$ (eV) & $\lambda$ (meV) & $\gamma$ (eV) & $\Gamma_i$ (meV) & $\Gamma_e$ (meV)  \\  \hline\hline
		MoS$_2$ & 3.18 \cite{Rasmussen2015} & 44.3 \cite{Olsen2016} & 2.5$^*$ & 14.4 \cite{Rasmussen2015} & 1.51 \cite{Rasmussen2015} & 25 \cite{Scuri2018} & 4/10/53 \cite{Cadiz2017}  \\ \hline
		WSe$_2$ & 3.32 \cite{Rasmussen2015} & 46.2 \cite{Olsen2016} & 2.38$^*$ & 47.2 \cite{Rasmussen2015} & 1.54 \cite{Rasmussen2015} & 32 \cite{Steinleitner2017} & 8/12/43 \cite{Christiansen2017}  \\ \hline
		\multicolumn{8}{l}{{\footnotesize $*$ We estimate the value of $\Delta$ by fitting the A and B resonances to the reported experimental data.}} \\
	\end{tabular}

\end{table}

\bibliography{SuppSHE}